\documentstyle[12pt,epsf]{article}

\newlength{\dinwidth}
\newlength{\dinmargin}
\begin{document}

\baselineskip15pt

\thispagestyle{empty}

\begin{flushright}
\begin{tabular}{l}
IASSNS-HEP-95/96\\
FTUOV-95/10\\
\end{tabular}
\end{flushright}

\vspace*{2cm}

{\vbox{\centerline{{\Large{\bf QUANTUM CORRECTIONS 
IN TWO-DIMENSIONAL}}}}}

\vspace{0.8cm}

{\vbox{\centerline{{\Large{\bf 
NON-SUPERSYMMETRIC HETEROTIC STRINGS
}}}}}

\vskip30pt

\centerline{M.\,A.\,R. Osorio\footnote{E-mail address:
    {\tt osorio@string1.ccu.uniovi.es}}}

\vskip6pt
\centerline{{\it Dept. de F\'{\i}sica, Universidad de Oviedo}}
\centerline{{\it Avda. Calvo Sotelo 18}}
\centerline{{\it E-33007 Oviedo, Spain}}

\vskip6pt

\centerline{{\it and}}

\vskip6pt

\centerline{M.\,A. V\'azquez-Mozo\footnote{E-mail address: {\tt
      vazquez@sns.ias.edu}}}

\vskip8pt
\centerline{{\it School of Natural Sciences}}\vskip2pt
\centerline{{\it Institute for Advanced Study}}\vskip2pt
\centerline{{\it Olden Lane, Princeton NJ 08540}}\vskip2pt
\centerline{{\it USA}}

\vskip .25in

\indent

\vspace*{24pt}

\noindent
We study quantum corrections  for a family of 24 non-supersymmetric heterotic
strings in two dimensions. We compute their genus two cosmological constant
using the hyperelliptic formalism and the genus one two-point functions for
the massless states. From here we get the mass corrections
to the states in the massless sector and discuss the role of the infrared
divergences that appear in the computation. We also study some tree-level
aspects of these theories and find that they are classified not only by
the corresponding Niemeier lattice but also by their {\it hidden} right-moving
gauge symmetry.

\vspace*{1.5cm}
\noindent
November, 1995

\setcounter{page}{0}
\setcounter{footnote}{0}

\newpage

\def\theequation{\thesection .\arabic{equation}}

\section{Introduction}
\label{int}
\setcounter{equation}{0}

It is well known that string theory is not just equivalent to a collection of
quantum fields. Atkin-Lehner symmetry is probably the best mathematical 
formulation of this physical statement as we have a net number of bosonic
massless fields not balanced with fermionic ones and 
at the same time the one-loop vacuum energy vanishes \cite{Moore}.
The vanishing of the cosmological constant due to
Atkin-Lehner symmetry is known to be a one loop effect and the common belief
is that interactions should actually induce a non-zero vacuum energy. 
Our objetive here will be to study these quantum corrections not only
to the model with Atkin-Lehner symmetry but also to all its 23 relatives
constructed on the Niemeier lattices.

Quantum corrections in heterotic string theories have been mainly studied in 
the 
context of supersymmetric models \cite{GSW,ADS,G,ADGN,Minahan2}. In these cases
non-renormalization theorems ensure that some loop 
correlation functions for less that four external massless particles have 
to vanish \cite{DP} since they are related to couplings in the low-energy 
action that are fixed by supersymmetry. However in the case of 
non-supersymmetric heterotic strings one expects 
to have an induced cosmological constant and/or a mass 
renormalization for the massless states.

Non-supersymmetric heterotic strings are peculiar for a number of reasons.
Among all the possible models that can be constructed 
only a relatively small subset gives raise to well-behaved string theories
in which the breaking of space-time supersymmetry does not introduce
tachyonic states. This is the case, for example, of the ten-dimensional
heterotic string with gauge group $SO(16)\times SO(16)$ \cite{LAG}; this
model is tachyon free and has a finite non-vanishing one loop cosmological
constant. However the theory is non-finite due to divergences appearing
in the computation of some loop amplitudes. More
recently the concept of {\it misaligned supersymmetry} \cite{Dienes}
has been introduced
in order to establish the minimum requirements a
non-supersymmetric heterotic string  has to fulfill 
in order to be tachyon free and then to have finite one-loop
induced vacuum energy. Nevertheless,
the computation of scattering amplitudes for these models may be
afflicted with infrared divergences as we will discuss 
in more detail later.

In the context of the model with Atkin-Lehner symmetry there arises an
obvious question which is how two-loops string physics modifies the
delicate cancellation of the one-loop vacuum energy. The best way to
answer this question is of course to compute its genus two cosmological
constant. This is not an easy task as
it  can be seen from the physics literature
during past years. The best way to accomplish such a computation
seems to be using hyperelliptic formalism which has provided good results
for supersymmetric heterotic strings \cite{knizhnik,tomas}. However
the expressions for the two-loops cosmological constant in a non-supersymmetric
model may, and will, be quite unmanageable. This is a little bit
disappointing but by no means the end of the story; as it was described
in \cite{moore-bos} in the context of the bosonic string, if we look at
the regions in the boundary of the genus two moduli space 
we can get some information about possible 
divergences. In particular, when a tubular neighborhood of a non-trivial
homology cycle gets long and skinny we can rewrite the contribution 
to the genus-two vacuum energy in terms of the on-shell two-point
function on the torus for the states in the model \cite{moore-bos}. 
In a sense we can partly satisfy our
curiosity about the physics that arise at two-loops by just looking
at some one-loop amplitudes.

The paper is organized as follows. In Sec. \ref{2} we describe
the two-dimensional models under study and
make a summary of some well-known and not-so-well-known facts about
them. In Sec. \ref{field} we study the underlying 
symmetries of the models and give an alternative contruction based
on free world-sheet fermions. We finish the section with the 
computation of tree-level correlation functions and the construction
of the low-energy effective action for the massless states.
In Sec. \ref{genus2} the 
genus two cosmological constant for these heterotic strings is
computed. Secs. \ref{nb}-\ref{nc} are devoted to the 
computation of the two-point functions on the torus for external 
massless states and in Sec. \ref{dis} we study the possibility
of calculating mass corrections from the gotten results. 
Finally in Sec. \ref{con} we will summarize the conclusions. For the 
sake of self-containment, some useful results
about lattice theta functions, Riemann surfaces in hyperelliptic formalism
and the Weierstrass elliptic function are presented in the Appendices.

\section{Heterotic Strings in Two-Dimensions}
\label{2}
\setcounter{equation}{0}

We are going to focus ourselves in the study of
a family of two-dimensional heterotic  strings \cite{Moore,ZfP,MM1,MM2}. 
These models are constructed by directly compactifying the
left-moving sector of a 26-dimensional bosonic string into one of the 24 
Niemeier lattices \cite{Lerche}. The right-movers are those of a type II 
string compactified using the $\Gamma_{8}$ lattice\footnote{i.e.,
the root lattice of $E_{8}$, the only even, self-dual lattice in eight 
dimensions.}. However, as they
stand, these 24 string models are supersymmetric. To break supersymmetry we
mod out the right moving sector by the operator \cite{Ginsparg-Vafa}
\begin{equation}
\alpha = (-1)^{F} e^{2\pi i P_{R}\cdot \delta},
\end{equation}
where $F$ is the target space fermion number,
$P_{R}$ is the momentum in the $\Gamma_{8}$ lattice and $\delta$
is a vector such that $2\delta \in \Gamma_{8}$. In our case we take
\begin{equation}
\delta = \left( \left(\frac{1}{2}\right)^{4},0^{4}\right).
\end{equation}
As it is usual in order to preserve modular invariance, in addition to the 
untwisted sector whose states are $\alpha$-invariant  
we must add up the twisted states in which the string closes modulo a 
transformation by $\alpha$ and then project again onto the states
invariant under this operator. At the end, we have to 
consider four subsectors, two of which belong to the untwisted sector
and correspond to the following pairing between the four conjugacy classes
of $SO(8)$ and certain set of vectors \cite{Ginsparg-Vafa}
\begin{eqnarray*}
(\Gamma_{8}^{+},v),
 \hspace*{2cm} (\Gamma_{8}^{-},s), 
\end{eqnarray*}
where $\Gamma^{\pm}_{8}$ are the subset of vectors in $\Gamma_{8}$ such
that their scalar product with $\delta$ is respectively an integer or a 
half-integer. In the twisted sector we have
\begin{eqnarray*}
(\Gamma_{8}^{+}+\delta,o),
\hspace{2cm} (\Gamma_{8}^{-}+\delta,c).
\end{eqnarray*}

It is easy to see what the massless spectrum for each of these models is.
Before the modding, the massless states are
\begin{eqnarray*}
\bar{\alpha}^{I}_{-1}|P_{L}^{2}=0\rangle \otimes |i\rangle , 
&\hspace*{2cm}&
\bar{\alpha}_{-1}^{I}|P_{L}^{2}=0\rangle \otimes |a\rangle , \nonumber
\\
|P_{L}^{2}=2\rangle \otimes |i\rangle ,
&\hspace*{2cm}&\hspace*{0.7cm}
|P_{L}^{2}=2\rangle \otimes |a\rangle ,
\end{eqnarray*}
where $\{|i\rangle,|a\rangle\}$ span the ${\bf 8_{v}}\oplus {\bf 8_{s}}$
of $Spin(8)$.
The states in the first line correspond to neutral particles under the
left-moving 
gauge group, while the particles in the second line are charged. All
the states have $P_{R}^{2}=0$.

When modding out by $\alpha$  none of the 
fermionic states in the massless sector survive the projection, since
they have $\alpha=-1$ so we are left with one-half of the states (those in
the first column). To construct the states in the twisted sector we go 
to the mass formula 
\begin{equation}
\alpha^{'}m^{2}_{R}=\frac{1}{2}P_{R}^{2}+\sum_{n>0} \alpha_{-n}^{i}
\alpha^{i}_{n}+\sum_{r>1/2} r S^{a}_{-r}S^{a}_{r} - \frac{1}{2},
\end{equation}
where $P_{R} \in \Gamma_{8}^{\pm}+\delta$ and $r$ runs over positive 
half-integers\footnote{The appearance of half-integers $r$ and the normal
ordering constant $-1/2$
are due to the
fact that in the twisted sector we have $S^{a}(\sigma+\pi)=(-1)^{F}S^{a}(
\sigma)=-S^{a}(\sigma)$.}.
The only way to get an $\alpha$-symmetric state with 
$m^{2}=0$ is to have $P^{2}_{R}=1$ and $N_{S}=N_{\alpha}=0$, which means that
it must belong to the $(\Gamma_{8}^{+}+\delta,o)$ sector. In fact it 
can be seen that there are $16$ such states corresponding to the $16$ points
in the $\Gamma_{8}^{+}+\delta$ lattice at $P_{R}^{2}=1$. Then we finally
find the following states in the massless
sector
\begin{eqnarray*}
\bar{\alpha}_{-1}^{I}|P_{L}^{2}=0\rangle \otimes |i\rangle ,&\hspace*{2cm}&
|P_{L}^{2}=2\rangle\otimes |i\rangle , \\
\bar{\alpha}_{-1}^{I}|P_{L}^{2}=0\rangle \otimes |P_{R}^{2}=1;0_{t}\rangle ,
&\hspace*{2cm}& |P_{L}^{2}=2\rangle\otimes |P_{R}^{2}=1;0_{t}\rangle ;
\end{eqnarray*}
$0_{t}$ in the right-moving part indicates the {\it twisted}
vacuum defined by $S_{r}^{a}|0\rangle=0$ with $r \geq 1/2$.
As a matter of fact we can divide these states into 
$24\times 24=576$ neutral bosonic 
particles plus $24\times r_{\Gamma}(1)$ charged bosons where $r_{\Gamma}(1)$
is the number of sites at $P_{L}^{2}=2$ in the corresponding Niemeier lattice.
A quite remarkable property of this family of models is that the spectrum
is Bose-Fermi degenerate in all mass levels except in the massless sector.
In fact we can rewrite the massless states using the
Neveu-Schwarz-Ramond (NSR) rather than Green-Schwarz (GS) formulation; this
will be useful later when constructing vertex operators for the 
massless states. The states in the untwisted sector in NSR language 
can be easily read from the ones in GS formulation. In the case of 
the twisted states one only has to take into account that the ground
state in the scalar conjugacy class of SO(8) is the standard NS vacuum
$|0_{NS}\rangle$, so we have
\begin{eqnarray*}
\bar{\alpha}^{I}_{-1}|P_{L}^{2}=
0\rangle\otimes b_{-\frac{1}{2}}^{i}|0_{NS}\rangle ,
&\hspace{2cm}& |P_{L}^{2}=2\rangle\otimes b_{-\frac{1}{2}}^{i}|0_{NS}\rangle ,
\\
\bar{\alpha}^{I}_{-1}|P_{L}^{2}=0\rangle\otimes |P_{R}^{2}=1;0_{NS}\rangle ,
&\hspace{2cm}& 
|P_{L}^{2}=2\rangle\otimes |P_{R}^{2}=1;0_{NS}\rangle .
\end{eqnarray*}

Two-dimensional heterotic strings can also be formulated using 
fermionic constructions \cite{Antoniadis,KLT,FKP}.  
In the fermionic model the right-moving sector is made out of a set of
24 free Majorana-Weyl fermions. In our case we will consider that all 
of them have the same boundary conditions on the world-sheet. In the 
path integral computation each sector of boundary conditions contributes
with a definite sign that is fixed by the requirement that the
resulting amplitude must be modular invariant. At one-loop level it can be
seen that the correct choices for the signs are
\begin{equation}
C(A,A)=-C(P,A)=-C(A,P)=1,
\label{p}
\end{equation}
where A and P stand for periodic or antiperiodic boundary conditions along
each of the two homology cycles of the torus. The computation of the 
partition function and other observables can be simplified by bosonizing
the fermions; then we are left with 12 free bosons living in the $D_{12}$ 
root lattice. The choice of signs implies that the only 
conjugacy classes that
contribute to the partition funcion are the vectorial and one of the 
spinorials.
The massless spectrum can then be constructed. To this purpose one can use
the familiar techniques used in the light-cone quantization of the superstring
to find 
\begin{equation}
\bar{\alpha}_{-1}^{I}|P_{L}^{2}=
0\rangle\otimes b_{-\frac{1}{2}}^{A}|NS\rangle ,
\hspace*{2cm} |P_{L}^{2}=2\rangle \otimes b_{-\frac{1}{2}}^{A}|NS\rangle ,
\end{equation}
with $A=1,\ldots,24$ and $|NS\rangle$ the Neveu-Schwarz vacuum
(not to be confused with $|0_{NS}\rangle$, the Nevew-Schwarz vacuum
in the bosonic construction of the right-moving sector).
It can be seen that the lowest state in the Ramond sector is in the 
first massive level \cite{Antoniadis}. Again, with $m^{2}=0$ we have
$24\times 24=576$ neutral states and $24\times r_{\Gamma}(1)$ charged
ones. One can wonder now whether or not the bosonic and fermionic
constructions give the same theory. In fact their partition functions
are equal, they have the same massless spectrum and it can be seen that the 
number of states in the vectorial+scalar of $SO(8)$ in the bosonic 
construction equals the number of states in the vectorial of $SO(24)$ in the
fermionic representation\footnote{Since both theories are Bose-Fermi degenerate
for $m>0$ the number of fermionic states are also the same.}. 
In the next section we will argue whether a fermionic construction exists
for the 24 heterotic models described earlier in a bosonic fashion.
To find such a construction will allow us to use either the bosonic
of the ferminonic formulation depending on what is the representation
in which the computation is simpler.

Among the menagerie of 24 models, either in the fermionic or 
the bosonic
realization, described above there is one whose properties deserve some
attention. This is the model which is built up using the Leech 
lattice; the main caracteristic of this lattice is the fact that  it has 
no points at (length)$^2$=2. This means that $r_{\Gamma}(1)=0$ and then
the massless spectrum is only made of neutral bosons. It
most unexpected property is that it has no
one-loop induced cosmological constant \cite{Moore,ZfP}. The vanishing
of the one loop vacuum energy is mathematically explained by the presence
of a discrete symmetry, called Atkin-Lehner symmetry, which
acts on the torus modular parameter $\tau$.  
Although this cancellation mechanism was for some time regarded as a
promising candidate to solve the cosmological constant problem, it was
soon realized \cite{balog-tuite} that this two-dimensional model was
essentially the only consistent theory with Atkin-Lehner symmetry, since
for any theory in higher dimensions the presence of this symmetry leads
to the existence of fermionic tachyons which is forbidden by Lorentz
invariance.   

One can ask for the 
physical meaning of this phenomenon. It is well known that in field theory
the only way to set the cosmological constant to zero without fine
tunning is by considering theories in which there is a Bose-Fermi
degeneration, i.e. supersymmetric theories\footnote{Witten has pointed 
out recently \cite{witten} that 
in (2+1)-dimensional supergravity, 
due to a conical singularity at infinity, there are no global supercharges
and then 
the vanishing of the cosmological constant could be 
accomplished without having Bose-Fermi degeneracy.}.
In this case however we
have a theory that while having a net number of bosonic massless states 
has no one-loop 
vacuum energy. In spite of the stringy nature of this cancellation
one would like to understand it in terms of field-theoretical degrees 
of freedom. The key to such an interpretation was given in \cite{MM1}. 
There the toroidal compactification of these models into ${\bf R}\times
S^{1}$ was studied and a non-analytic behavior of the partition
function as a function of the compactification scale was found
 at the self-dual 
size. It was also shown that the part of the partition 
function below the self-dual radius contains the contribution of $24\times 24$ 
bosonic states with the wrong sign, while above $\sqrt{\alpha^{'}}$
it contains a constant term which is equal to $-24\times 24$ times the vacuum 
energy
of the $c=1$ model. This exactly cancels the contribution to the vacuum 
energy of the net bosonic states in the Atkin-Lehner model as computed 
using the analog model. 

The moral of the story is that if we want to understand the zero of
the cosmological constant in the model with Atkin-Lehner symmetry 
in terms of field theory we need to introduce some {\it intruder}
states which contribute to the partition function with the wrong 
sign and that when compactifying one of the open dimensions only
get excited below the Planck scale. In the decompactification
limit their contribution is just given by 
\begin{equation}
\Lambda_{intruder}=24\times 24\int_{{\cal F}}\frac{d^{2}\tau}{\tau_{2}^{2}}
\end{equation}
and then it cancels exactly the regularized vacumm energy of the massless
states in the Atkin-Lehner model\footnote{
We have defined the vacuum energy as minus the integral to the fundamental
region of the partition function in such a way that bosonic states contribute
with a minus sign.}. 
Of course this by no means implies
the real existence of these states. They are just the result of 
trying to explain a stringy phenomenon using field-theoretical words. 

Let us look back to the partition function in the
bosonic construction. The right moving part is 
\begin{equation}
Z_{R}=\frac{\theta_{3}^{4}-\theta_{4}^{4}}{2\eta^{12}}\Theta_{\Gamma_{8}^{+}}
-\frac{\theta_{2}^{4}}{2\eta^{12}}\Theta_{\Gamma^{-}_{8}}+
\frac{\theta_{3}^{4}+\theta_{4}^{4}}{2\eta^{12}}\Theta_{\Gamma_{8}^{+}
+\delta}-\frac{\theta_{2}^{4}}{2\eta^{12}}\Theta_{\Gamma_{8}^{-}+\delta}.
\end{equation}
In fact, the theta functions associated with the four sets of vectors can
be rewritten in terms of the theta function for $\Gamma_{8}$ with 
characteristics
\begin{eqnarray}
\Theta_{\Gamma_{8}^{\pm}}&=&\frac{1}{2}\Theta_{\Gamma_{8}}\left[
\begin{array}{c}
0 \\
0
\end{array}
\right] \pm \frac{1}{2}
\Theta_{\Gamma_{8}}\left[
\begin{array}{c}
0 \\
\delta
\end{array}
\right], \nonumber \\
\Theta_{\Gamma_{8}^{\pm}+\delta}&=& \frac{1}{2}\Theta_{\Gamma_{8}}
\left[
\begin{array}{c}
\delta \\
0
\end{array}
\right]
\pm \frac{1}{2}\Theta_{\Gamma_{8}}\left[
\begin{array}{c}
\delta \\
\delta
\end{array}
\right],
\end{eqnarray}
so the contribution of the right-moving modes, $Z_{R}$, to the  
partition function can be rewritten as
\begin{eqnarray}
Z_{R}&=&\frac{\theta_{3}^{4}-\theta_{4}^{4}-\theta_{2}^{4}}{4\eta^{12}}
\Theta\left[
\begin{array}{c}
0 \\
0
\end{array}
\right]+\frac{\theta_{3}^{4}-\theta_{4}^{4}+\theta_{2}^{4}}{4\eta^{12}}
\Theta\left[
\begin{array}{c}
0 \\
\delta
\end{array}
\right] \nonumber \\
&+& \frac{\theta_{3}^{4}+\theta_{4}^{4}-\theta_{2}^{4}}{4\eta^{12}}
\Theta\left[
\begin{array}{c}
\delta \\
0
\end{array}
\right]+\frac{\theta_{3}^{4}+\theta_{4}^{4}+\theta_{2}^{4}}{4\eta^{12}}
\Theta\left[
\begin{array}{c}
\delta \\
\delta 
\end{array}
\right].
\label{orbifold}
\end{eqnarray}
Using this expression it is clear the orbifold-like structure of the
partition function. Since its left-moving part  
is not affected by the operator $\alpha$ we can write
\begin{equation}
Z= \sum_{m,n=0}^{1} Z_{L}(\bar{\tau})Z_{R}^{(m,n)}(\tau),
\label{sum}
\end{equation}
where $Z_{R}^{(m,n)}$ is the right moving contribution for 
the string with boundary conditions for the bosons 
twisted by $(\alpha^{m},
\alpha^{n})$ along the two homology cycles of the torus.
Let us notice that the first term in (\ref{orbifold})
is equal to zero because of Jacobi's {\it aequatio}.
Individually, each term in (\ref{sum}) can be written as a 
sum over spin structures $e$ with given phases $C_{e}(m,n)$ which can
be read from (\ref{orbifold})
\begin{equation}
Z_{R}^{(m,n)}=\frac{1}{4}
\sum_{e} C_{e}(m,n) \frac{\theta^{4} [e]}{\eta^{12}}\Theta_{(m,n)}(0|\tau);
\end{equation}
from now on, in order to simplify the expressions, we will write
\begin{equation}
\Theta_{(m,n)}(0|\tau)=\Theta\left[
\begin{array}{c}
m\delta \\
n\delta
\end{array}
\right](0|\tau).
\end{equation} 
There is no contribution
coming from world-sheet fermions with space-time indices, since this is 
cancelled by the contribution of the conformal and superconformal ghosts.  
The theta functions can be computed with the result
\begin{eqnarray}
\Theta_{(0,0)}&=& \frac{1}{2}(\theta_{3}^{8}+\theta_{4}^{8}+\theta_{2}^{8}), 
\nonumber \\
\Theta_{(1,0)}&=& \theta_{3}^{4}\theta_{2}^{4}, \nonumber \\
\Theta_{(0,1)}&=& \theta_{3}^{4}\theta_{4}^{4}, \nonumber \\
\Theta_{(1,1)}&=& \theta_{2}^{4}\theta_{4}^{4}.
\end{eqnarray}
On the other hand the left-moving partition function, which is common to 
all the sectors, can be written in terms of the modular invariant
function $j(\tau)$ as
\begin{equation}
Z_{L}(\bar{\tau})= \overline{j(\tau)}-720+r_{\Gamma}(1).
\end{equation}

\section{Fermionic Constructions, Gauge Symmetry
and the Low Energy Field Theory.}
\label{field}
\setcounter{equation}{0}

In the previous section we have discussed the construction of
two-dimensional heterotic string models without space-time
supersymmetry. Now we are going to study more carefully the
fermionic realization of the family of 24 heterotic strings.
After doing this we will try to extract the effective low-energy field
theory for the massless particles.

At first sight there is an obvious asymmetry between the massless
sectors in the bosonic (B) and fermionic (F) construction. In the B models 
massless particles 
are of two very different types\footnote{
In what follows we will drop the left-moving parts of the
states whenever they are not relevant for the discussion.}; on one hand
we have the 8 untwisted states $|i;P_{R}=0\rangle$ which are in 
the vector of $SO(8)$ and on the other we find the 16 twisted
states $|P_{R}^{2}=1\rangle$ associated with the 16 vectors
in $\Gamma_{8}^{+}+\delta$ with $P_{R}^{2}=1$. On the contrary in 
the F model we are left with 24 states $b_{-1/2}^{A}|NS\rangle$ in the
vector of $SO(24)$. It is not very pleasant to have such an 
asymmetry when we would like to identify both models.

In order to solve the mistery, let us look more closely to the B model.
The 16 possible vectors $P_{R}$ have coordinates in the orthonormal basis
\begin{equation}
\left(\left(\pm\frac{1}{2}\right)^{4},0^{4}\right), \hspace*{2cm}
\left(0^{4},\left(\pm\frac{1}{2}\right)^{4}\right),
\label{vectors}
\end{equation}
with an even number of minus signs. In fact the 16 vectors (\ref{vectors})
can be ordered in 8 pairs $\{ P_{R}^{(i)},-P_{R}^{(i)}\}$, $(i=1,\ldots,8$)
such that $P_{R}^{(i)}\cdot P_{R}^{(j)}=0$ when $i\neq j$. One can 
easily see that the set (\ref{vectors}) is isomorphic to the root system
of $SU(2)^{8}\simeq SO(3)^{8}$ (modulo a rescaling of the roots). Morover, 
the 8
states in the untwisted sector fill the states in the Cartan subalgebra of
the same Lie algebra. With this result at hand the most we can say
is that the states in the massless sector in the B construction fit in
the adjoint representation of $SU(2)^{8}$. However in order to show that
this is realized as a gauge symmetry of the theory we have to give a 
step forward and prove that there is a realization of the current 
algebra of $SU(2)^{8}$ in the algebra of vertex operators. 

Using the NSR formulation, the right-moving parts of the vertex operators in 
the 0 picture for states 
in the massless sector are
\begin{eqnarray}
V_{0}^{j}(k,z)&=&[\partial_{z} X^{j}+i(k_{\mu}\psi^{\mu})\psi^{j}]e^{ik_{\mu}
X^{\mu}(z)}, \nonumber \\
V_{0}^{P_{R}}(k,z)&=& iP_{R}^{j}\psi^{j}e^{iP_{R}^{i}X^{i}(z)+ik_{\mu}X^{\mu}
(z)}.
\end{eqnarray}
The second expression can be easily obtained by taking into account
that the oscillator part of the twisted state is just that of the NS
vacuum\footnote{In the GS formulation, the contruction of the vertex operator
for the twisted states requires the introduction of a new field 
$\sigma^{i}(z)$, ($i=1,\ldots,8$)
which creates the $|i\rangle$ vacuum out of
the twisted vacuum, $|i\rangle=\sigma^{i}(0)|0_{t}\rangle$.}. 
If we compute the OPE of these vertex operators at $k=0$ we
find
\begin{eqnarray}
V^{P_{R}}_{0}(z)V^{-P_{R}}_{0}(w)
&=&\frac{1}{(z-w)^{2}}-\frac{1}{z-w}iP_{R}^{i}V^{i}_{0}(w),
\nonumber \\
V^{i}_{0}(z)V_{0}^{P_{R}}(w)
&=& \frac{-i}{z-w} P_{R}^{i}V^{P_{R}}_{0}(w), \nonumber \\
V^{i}_{0}(z)V^{j}_{0}(w)&=&\frac{-\delta^{ij}}{(z-w)^{2}},
\label{km}
\end{eqnarray}
with $V^{P_{R}}(z)V^{P_{R}^{'}}(w)=0$ when $P_{R}^{'}\neq -P_{R}$. At this
point after rescaling $V^{j}\rightarrow iV^{j}$ one 
is tempted to identify (\ref{km}) as the OPE corresponding to 
the $k=1$ $SU(2)^{8}$ Ka\u{c}-Moody algebra in the Cartan-Weyl basis. However
this would not be correct; the reason is that $P_{R}^{2}=1$, contrary to
the $SU(2)^{8}$ roots which are canonically normalized to $\alpha^{2}=2$.
This is relevant, since the components of the roots correspond to the
structure constants of the Lie algebra in the Cartan-Weyl basis. In order to
recover the standard form of the $\widehat{SU(2)^{8}}$ 
affine algebra we have
to renormalize the vertex operators as $V_{0}\rightarrow \sqrt{2}V_{0}$.
After this we get the canonical OPE of a level 2 $SU(2)^{8}$ 
Ka\u{c}-Moody
algebra. Then we see that
the introduction of the twisted states enhances the right-moving gauge 
symmetry from $U(1)^{8}$ to $SU(2)^{8}$ and then the full 
symmetry of the string theory is $G_{L}\times SU(2)^{8}$ with $G_{L}$ the
gauge group associated with the corresponding Niemeier lattice 
(or $G_{L}=U(1)^{24}$ for the Leech lattice \cite{Lerche}).

Let us move to the F models. Now the internal CFT is that of 
a system of 24 Majorana-Weyl fermions all of them with the same 
world-sheet boundary conditions. In such a system there is a
$N=1$ superconformal symmetry generated by non-linear transformations
\cite{Antoniadis}
\begin{equation}
\delta_{\epsilon}\lambda^{A}= \frac{i\epsilon}{\sqrt{2 C_{2}
({\cal G})}}f^{ABC}\lambda^{B}\lambda^{C},
\end{equation}
where $f^{ABC}$ are the structure constants of a semisimple Lie algebra
${\cal G}$ (${\rm dim\,}{\cal G}=24$), $C_{2}({\cal G})$ is the quadratic 
Casimir of the adjoint 
representation of ${\cal G}$
and $\epsilon$ is an anticommuting infinitesimal parameter. Combining the 
fermions $\lambda^{a}$ with
the structure constants one can construct the following currents
\begin{equation}
J^{A}(z)=\frac{i}{2}f^{ABC}\lambda^{B}\lambda^{C},
\end{equation}
which generate an affine algebra $\widehat{\cal G}$ with level
$k=C_{2}({\cal G})/2$. 
In fact it can be shown \cite{Antoniadis}
that all fermionic models can be classified in terms of a pair of semi-simple
Lie groups $G$, $H$ such that $H\subset G$ and $G/H$ is a symmetric space
(${\cal G} = {\rm Lie\,}(G)$) \cite{barut}\footnote{$G/H$ is a 
symmetric space if there exists an involutive autormorphism $\sigma$
in $G$ such
that $G_{\sigma}^{0}\subset H\subset G_{\sigma}$, where $G_{\sigma}$
is the set of points in $G$ fixed by $\sigma$ and $G_{\sigma}^{0}$ is
its identity component.}. This last condition implies that 
the theory can be truncated without breaking $N=1$ superconformal symmetry
by projecting on the states with 
$(-1)^{F_{pseudo}}=1$,
where $F_{pseudo}$ is the fermion number for the $\lambda^{i}$
with indices in ${\cal G}-{\cal H}$.
This modding breaks the actual gauge symmetry of the system 
from $G$ to $H$.

To connect with the B model we only have to take $G=H=SU(2)^{8}$ so
we take the same GSO projection for all worldsheet fermions. The
right moving part of the 
vertex operators for the massless states in the 0 picture
are
\begin{equation}
V^{A}_{0}(k,z)= \left[ \frac{i}{\sqrt{2}}f^{ABC}\lambda^{B}\lambda^{C}+
i(k_{\mu}\psi^{\mu})\lambda^{A}\right] e^{ik_{\mu}X^{\mu}(z)}.
\end{equation}
These vertex operators create states that are
in the adjoint representation of $SU(2)^{8}$ and that when taken at 
zero external momentum generate a
$\widehat{SU(2)^{8}}_{k=2}\simeq \widehat{SO(3)^{8}}_{k=2}$ 
Ka\u{c}-Moody algebra.

Then we have two constructions, bosonic and fermionic, with the same
underlying symmetry,
namely $\widehat{SU(2)}^{8}_{k=2}$. In fact the two constructions (B and F)
give the same answers
when computing scattering amplitudes as can be checked at
tree level (genus zero).
In forthcoming sections 
we will also see that this is true in one-loop calculations. 
This strongly suggests that the B and F constructions render
string theories that are completely equivalent. 

To finish this section we are going to get the low-energy
effective theory for the on-shell massless states. As we shaw  these 
states are 192 neutral untwisted bosons $\Phi^{Ii}$ ($I=1,\ldots,
24$; 
$i=1,\ldots,8$), 384 neutral twisted ones $\Psi^{I,P_{R}}$ ($P_{R}$ runs
over all $P_{R}\in\Gamma_{8}^{+}+\delta$ with $P_{R}^{2}=1$) and the 
corresponding charged particles, $8\times r_{\Gamma}(1)$
$\Phi^{\alpha,i}$ and $16\times r_{\Gamma}(1)$ $\Psi^{\alpha,P_{R}}$ in
the untwisted and twisted sectors respectively, 
with $\alpha$ running over the roots of the left-moving 
gauge group. However from our previous discussion we know that
the massless states are in the $(adj,adj)$ representation of the 
gauge group $G_{L}\times SU(2)^{8}$. Then we can write them in
shorthand as a single field $\Phi=\Phi^{\bar{A}A}T^{\bar{A}}_{L} 
\otimes T^{A}_{R}$ 
where $T_{L}^{\bar{A}}$ and $T_{R}^{A}$ are the generators 
in the fundamental representation of
$G_{L}$ and $SU(2)^{8}$ respectively.

To get the couplings between the low-energy fields
we have to compute the scattering amplitudes for the corresponding 
vertex operators at tree level in the string loop expansion. 
Two point functions vanish, reflecting the fact that the string equations
are satisfied at tree level. The coupling involving three fields can be
easily computed using either the B of F model 
with the resulting term in the effective action
\begin{equation}
A_{3}=\frac{1}{\sqrt{2}}
f_{L}^{\bar{A}\bar{B}\bar{C}}f_{R}^{ABC}\Phi^{\bar{A}A}\Phi^{\bar{B}B}
\Phi^{\bar{C}C}
=\frac{1}{\sqrt{2}}{\rm Tr\,}\{[\Phi,\Phi]\Phi
\},
\end{equation}
where we have introduced the left and right-moving structure constants
and the commutator has to be understood as the tensor product of  
commutators for the left and right-moving generators.
Because of the presence of $f_{L}^{\bar{A}\bar{B}\bar{C}}$ we see that this 
coupling vanishes
for the theory constructed in the Leech lattice, in which the left-moving
group is abelian. In the general case in which $r_{\Gamma}(1)\neq 0$ the
coupling  exists but only between one untwisted and two twisted
states with opposite values of $P_{R}$ (or in other words, between one
neutral and two charged states with total $SU(2)^{8}$-charge equal to zero).
This can be understood from the known results in lower-dimensional 
heterotic strings \cite{lust-theisen}: the right-moving part of the 
amplitude for three gauge bosons is given by the contraction of the 
polarization tensors with space-time momenta $\zeta^{\mu}_{i}k_{j\mu}$. 
For untwisted states, polarizations lie always in the internal space and
thus are orthogonal to all space-time momenta, forcing the amplitude to
vanish. On the contrary when twisted states are present we have internal
momenta $P_{R}$ and then the right-moving part of the amplitude does not 
vanish but it is proportional to the $P_{R}^{i}$ which are essentially the
structure constants for $SU(2)^{8}$ in the Cartan-Weyl basis.

In the case of the four-fields coupling the computation is a little
more involved since a Koba-Nielsen integral has to be performed. Taking
the leading terms in the limit $\alpha^{'}\rightarrow 0$ it can be seen
that the corresponding contribution to the low-energy action is
\begin{eqnarray}
A_{4}&=&\frac{1}{2}f_{L}^{\bar{A}\bar{B}\bar{E}}f_{L}^{\bar{E}\bar{C}\bar{D}} 
f_{R}^{ABE}f_{R}^{ECD}
\Phi^{\bar{A}A}\Phi^{\bar{B} B}\Phi^{\bar{C} C}\Phi^{\bar{D} D}
+\frac{\alpha^{'}}{8}\left(
f_{L}^{\bar{A} \bar{B} \bar{E}}f_{L}^{\bar{A} \bar{B} \bar{E}} \delta^{AC}
\delta^{BD} \right.\nonumber \\
&+&\frac{1}{2}\left.\delta^{\bar{A} \bar{C}}\delta^{\bar{B} \bar{D}}
f_{R}^{ABE}f_{R}^{ECD}\right)
\partial_{\mu}\Phi^{\bar{A} A}\Phi^{\bar{B} B}\partial^{\mu}\Phi^{\bar{C} C}
\Phi^{\bar{D} D}
+O(\alpha^{'\,2}).
\label{4}
\end{eqnarray}

Now we can construct the low-energy field theory for the massless 
fields $\Phi$. Retaining only the leading terms in the $\alpha^{'}$ 
expansion the result is
\begin{eqnarray}
S&=&\frac{1}{2}\int d^{2}x\, {\rm Tr\,}\left\{\partial_{\mu}\Phi\partial^{\mu}
\Phi+\frac{g}{\alpha^{'}}[\Phi,\Phi]\Phi+\frac{g^{2}}{\alpha^{'}}
[\Phi,\Phi]^{2}\right\},
\label{action}
\end{eqnarray}
where $g$ is the dimensionless gauge coupling constant which is proportional
to the string coupling constant and inversely proportional to the 
square root of the product of the levels of the right and left-moving
Ka\u{c}-Moody algebras.
The effective action get simpler if we particularize
to the case of the Leech lattice since now all the commutators vanish
\begin{equation}
S_{Leech}=\frac{1}{2}\int d^{2}x\, {\rm Tr\,} 
\partial_{\mu}\Phi\partial^{\mu}\Phi\,,
\end{equation}
i.e., we are left with a sigma-model defined on 
$U(1)^{24}\otimes SU(2)^{8}$. 

\section{Genus-Two Cosmological Constant for the
\newline Two-Dimensional Models}
\label{genus2}
\setcounter{equation}{0}

One way to study the physics that 
arises after turning on the interaction between strings is to compute
the genus two vacuum energy. Higher genus computations
in string theory have been source of discussion along the years. In the 
ten-dimensional $E_{8}\times E_{8}$ or $SO(32)$
heterotic string some 
expressions have been proposed which vanish, as it is expected from 
supersymmetry 
\cite{knizhnik,morozov}. However the main drawback of these computations
is the fact that the vanishing expressions are not modular invariant.
In ref. \cite{tomas} a way of computing a two-loop (vanishing) modular
invariant cosmological constant was finally designed. In the case
of the supersymmetric heterotic string the two-loops cosmological constant
can be written as an integral over the fundamental region of $Sp(2,{\bf Z})$
of an expression which is identically zero due to some combinations
of standard 
Riemann identities. In our case, however, we do not expect this to be
the case and therefore the usual argument in favor of the expressions 
given in \cite{knizhnik,morozov} (that zero is always modular invariant) 
cannot even be applied. 
In the following computation we will closely follow ref. 
\cite{tomas}, which we regard as the most clarifying approach,
and we will be able to get a modular invariant expression for 
the integrand of the cosmological constant. We will use the fermionic
construction of the model in which the computations notably simplify.

The starting point is a modification of the Knihznik formula \cite{knizhnik}
for the two-loops cosmological constant in hyperelliptic formalism
(for definitions and notations see Appendix B)
\begin{equation}
Z_{g=2}=\sum_{e} C(e)\int \prod_{i=1}^{6} d^{2}a_{i} \frac{1}{dv_{pr}^{2}}
T^{-1}\overline{\prod_{k<l}a_{kl}^{-3}}\overline{F(\Lambda_{24})}
\prod_{k<l}a_{kl}^{-2}[{\cal P}^{X}+{\cal P}^{gh}_{e}]{\cal O}_{e}^{3};
\label{k2}
\end{equation} 
$F(\Lambda_{24})$
is the partition function for the left-moving bosonic sector 
and $C(e)$ are the phases that generalize (\ref{p}) for the right-moving
world-sheet fermions at genus two. The correlation of 
the two PCOs now is
a little bit different from the one for the ten-dimensional heterotic string
since a PCO has a space-time and an internal part $P_{+1}=P_{+1}^{s-t}+
P_{+1}^{int}$. The correlator then is written
\begin{equation}
\langle P_{+1}(z)P_{+1}(w)\rangle =
\langle P_{+1}^{s-t}(z)P_{+1}^{s-t}(w)\rangle +
\langle P_{+1}^{int}(z)P_{+1}^{int}(w)\rangle .
\end{equation}
Nonetheless, as we take the limit $z\rightarrow a_{1}$, $w\rightarrow a_{2}$
it can be seen that the internal part does not  contribute to (\ref{k2}) 
so we have
\begin{equation}
{\cal P}^{X} = \frac{1}{5}{\cal P}^{X}_{10},
\end{equation}
where ${\cal P}^{X}_{10}$ is given by (\ref{P10}).
For the ghost part we find just the same result since the ghost content
of the two-dimensional models is the same than in the ten-dimensional
case.

As it is argued in Appendix B, it is convenient to eliminate the 
$SL(2,{\bf C})$ redundancy by introducing the harmonic ratios (\ref{h}). 
Then (\ref{k2}) can be rewritten in terms of $\lambda_{i}$ as
\begin{equation}
Z_{g=2}=\sum_{e} C(e)\int \wedge_{i=1}^{3}d\lambda_{i}{\cal W}_{2}^{-1}(
\lambda_{i})
 \overline{{\cal F}
(\Lambda_{24})}[U^{X}_{e}+U_{e}^{gh}]
\label{k22},
\end{equation}
where
\begin{equation}
{\cal F}(\Lambda_{24})=\wedge_{i=1}^{3}d\lambda_{i}\frac{a_{12}a_{15}^{2}
a_{25}^{2}a_{35}^{2}a_{36}a_{46}}{a_{45}a_{56}}\prod_{k<l}a_{kl}^{-3}
\Theta(\Lambda_{24}),
\end{equation}
${\cal W}_{2}(\lambda_{i})=|a_{12}a_{45}a_{36}|^{2} T$
and
\begin{eqnarray}
U_{e}^{X}&=&\frac{1}{8}\prod_{k<l}a^{-2}_{kl}\frac{a_{12}a_{15}^{2}a_{25}^{2}
a_{35}^{2}a_{36}a_{46}^{4}}{a_{12}a_{45}a_{56}^{2}}\left\{
a_{23}a_{24}a_{25}\left(\frac{a_{16}^{2}}{a_{26}}\right){\cal P}_{12}+
(a_{1}\leftrightarrow a_{2})\right\}{\cal O}_{e}, \nonumber \\
U_{e}^{gh} &=& \prod_{k<l} a_{kl}^{-2}
\frac{a_{12}a_{15}^{2}a_{25}^{2}a_{35}^{2}
a_{36}a_{46}^{4}}{a_{45}a_{56}^{2}} {\cal P}_{e}^{gh}{\cal O}_{e}
\end{eqnarray}
and ${\cal P}_{12}$ is defined as
\begin{equation}
{\cal P}_{12}=\left(\frac{a_{26}}{a_{16}}\right)^{2}\frac{P_{12}}{T}.
\end{equation}

The strategy now is that of ref. \cite{tomas}. $Z_{g=2}$ as given by 
(\ref{k22}) is of the form
\begin{equation}
Z_{g=2}=\sum_{e} I_{(e)}.
\end{equation}
However, in general, the contributions $I_{(e)}$ are not even invariant 
under the 
subgroup of modular transformations $\Gamma_{e}$ that leaves the spin 
structure $e$ unchanged. Nevertheless not everything is lost since 
$I_{e}$ is invariant under a subgroup of $\Gamma^{'}\subset\Gamma_{e}$. 
Then we
can use the results of \cite{AOnp} and perform a coset extension from
$\Gamma^{'}$ to $\Gamma_{e}$. Once we have such a $\Gamma_{e}$-invariant
extension $\tilde{I}_{e}$ of $I_{e}$ we can further extend it to the 
full modular group by the same procedure. 

The final result of 
the coset extension to the full modular group will still  depend on the 
spin structure we started with \cite{tomas}. 
In fact we have two orbits of spin
structure contributions $I_{(e)}$ which cannot be transformed into
one another using modular transformations; these are the orbits that contain
respectively the contributions of the spin structures $(12A_{3}^{e}||
B_{1}^{e}B_{2}^{e}B_{3}^{e})$ and $(1A_{2}^{e}A_{3}^{e}||2B_{2}^{e}B_{3}^{e})$.
The way to decide between the two possible results is that the final
expression has to have
 the good factorization properties. In the ten-dimensional
heterotic string, and in our case also, the correct answer is gotten
by starting with the $(123||456)$ spin structure.

Then let us begin with $I_{(123||456)}$. The permutations of the branch
points that generate the subgroup $\Gamma_{(123||456)}
\subset \Gamma$ are $(12)$, $(23)$, $(45)$, $(56)$ and $(14)(25)(36)$. 
$I_{(123||456)}$ as read from (\ref{k22}) can be decomposed into a matter
part and a ghost part $I_{(123||456)}=I^{X}_{1}+I^{gh}_{1}$. Furthermore,
looking at the explicit expressions of $I^{X}_{1}$ and $I^{gh}_{1}$ we see
that they can be written respectively as
\begin{eqnarray}
I_{1}^{X}&=&I_{1,1}^{X}+(12)I_{1,1}^{X}, \nonumber \\
I_{1}^{gh}&=& I_{1,1}^{gh}+(45)I^{gh}_{1,1}+(56)(45)I_{1,1}^{gh},
\end{eqnarray}
where 
\begin{eqnarray}
I^{X}_{1,1} &=& \frac{1}{8}(\wedge_{i=1}^{3}d\lambda_{i}){\cal W}_{2}^{-1}
\overline{{\cal F}(\Lambda_{24})}\frac{(\lambda_{2}-\lambda_{1})(\lambda_{3}
-\lambda_{1})(\lambda_{3}-\lambda_{2})^{2}}{\lambda_{1}^{2}\lambda_{2}
\lambda_{3}^{2}(\lambda_{2}-1)^{3}(\lambda_{3}-1)}{\cal P}_{12}, \nonumber \\
I^{gh}_{1,1} &=& \frac{1}{4}(\wedge_{i=1}^{3}d\lambda_{i}){\cal W}_{2}^{-1}
\overline{{\cal F}(\Lambda_{24})}\frac{(\lambda_{2}-\lambda_{1})
(\lambda_{3}-\lambda_{1})^{2}(\lambda_{3}-\lambda_{2})}{\lambda_{1}
\lambda_{2}^{2}\lambda_{3}^{2}(\lambda_{1}-1)^{2}(\lambda_{2}-1)
(\lambda_{3}-1)}.
\label{I}
\end{eqnarray}
Then, since $I_{1,2}$ as well as $I^{gh}_{1,2}$ and $I^{gh}_{1,3}$ are obtained
from $I_{1,1}^{X}$ and $I_{1,1}^{gh}$ by transformations that belong
to $\Gamma_{(123||456)}$ we can make the coset extension directly from
(\ref{I}). 

Looking at $I^{gh}_{1,1}$ and applying the generators of 
$\Gamma_{(123||456)}$ we find
\begin{eqnarray}
(12)I_{1,1}^{gh} &=& -\frac{\lambda_{2}(\lambda_{3}-\lambda_{2})
(\lambda_{1}-1)}{\lambda_{1}(\lambda_{2}-1)(\lambda_{3}-\lambda_{1})}
I_{1,1}^{gh}, \nonumber \\
(23)I_{1,1}^{gh} &=& \frac{(\lambda_{2}-\lambda_{1})^{2}(\lambda_{3}-1)}{
(\lambda_{3}-\lambda_{1})^{2}(\lambda_{2}-1)}I_{1,1}^{gh}, \nonumber \\
(45)I_{1,1}^{gh} &=& \frac{\lambda_{2}}{\lambda_{1}}I^{gh}_{1,1}, \nonumber \\
(56)I_{1,1}^{gh} &=& I_{1,1}^{gh}, \nonumber \\
(14)(25)(36)I_{1,1}^{gh}&=&-\frac{\lambda_{2}-\lambda_{1}}{(\lambda_{3}
-\lambda_{1})(\lambda_{2}-1)}I_{1,1}^{gh}.
\label{Is}
\end{eqnarray}
These are exactly the same transformations that one finds for the case
of the ten-dimensional supersymmetric heterotic string. Following
the same steps than in \cite{tomas} we can see that after performing the
coset extension to the full modular group we are going to have 
$I^{gh}_{(1)}=0$. In fact the vanishing of the ghost contribution can
be seen in a more general context. The ghost part of the correlation
of the two PCOs, ${\cal P}_{e}^{gh}$, is a holomorphic function as can be
easily seen from its expression (\ref{gh1}) or (\ref{gh2}). This means 
that the integrand of $I^{gh}_{(1)}$ factorizes into a holomorphic and
an antiholomorphic function of the period matrix
\begin{equation}
\int I^{gh}_{(1)}=\int \prod_{i<j}^{2} d^{2}\tau_{ij} 
(\det{{\rm Im\,}\tau})^{-1}
(\overline{\Delta_{(2)}})^{-2}\,\overline{\Theta(\Lambda_{24})}
\,Z_{R}^{gh}(\tau)\,,
\end{equation}
where $\Delta_{2}=\prod_{e}\theta[e](0|\tau)$, the product being over the ten
even spin structures.
Since by construction $I^{gh}$ is modular invariant, $Z_{R}^{gh}$ must be 
a modular function of weight 2. Moreover if the theory has no right-moving
tachyons (as it is the case for both the supersymmetric heterotic string
and the two dimensional models under consideration)
$Z_{R}^{gh}$ must be not only a function but a weight 2 modular form under
$Sp(2,{\bf Z})$. However, as proved by Igusa \cite{igusa}, there is no 
modular functions of weight 2 at genus two, and then $Z_{R}^{gh}(\tau)=0$
(cf. \cite{moore-et-al}). 

Let us turn now to the matter part. 
Now the generators of $\Gamma_{(123||456)}$
act on $I^{X}_{1,1}$ as
\begin{eqnarray}
(12)I_{1,1}^{X}&=& -\frac{\lambda_{1}(\lambda_{3}-\lambda_{1})(\lambda_{2}
-1)^{3}}{\lambda_{2}(\lambda_{3}-\lambda_{2})(\lambda_{1}-1)^{3}}
\frac{{\cal P}_{21}}{{\cal P}_{12}} I^{X}_{1,1} = M_{(1),21}I_{1,1}^{X},
\nonumber \\
(23)I_{1,1}^{X}&=& -\frac{\lambda_{3}(\lambda_{2}-\lambda_{1})(\lambda_{2}
-1)}{\lambda_{2}(\lambda_{3}-\lambda_{1})(\lambda_{3}-1)}
\frac{{\cal P}_{13}}{{\cal P}_{12}} I^{X}_{1,1}= M_{(1),13}I^{X}_{1,1},
\nonumber \\
(45)I_{1,1}^{X} &=& I_{1,1}^{X}, \nonumber \\
(56)I_{1,1}^{X} &=& I_{1,1}^{X}, \nonumber \\
(14)(25)(36) I^{X}_{1,1} &=& \frac{
\lambda_{3}^{2}(\lambda_{2}-\lambda_{1})(\lambda_{2}
-1)}{\lambda_{2}(\lambda_{3}-\lambda_{2})(\lambda_{1}-1)^{2}}
\frac{{\cal P}_{45}}{{\cal P}_{12}} I^{X}_{1,1}
=M_{(1),45}I^{X}_{1,1},
\label{tI} 
\end{eqnarray}
where ${\cal P}_{ij}$ are the obvious generalizations of ${\cal P}_{12}$.
In general given a transformation $g\in\Gamma_{(123||456)}$ which takes
the pair $\{12\}$ to $\{ij\}$ we will have $g\cdot I_{1,1}^{X}= M_{(1),ij} 
I^{X}_{1,1}$ with $M_{(1),ij}$ proportional to ${\cal P}_{ij}/{\cal P}_{12}$
($M_{(1)12}=1$). 
If we want to avoid overcounting we have to consider only
one transformation $g$ such that $g(12)=(ij)$. This leaves  only
12 transformations and the $\Gamma_{(123||456)}$-invariant extension of
$I_{1,1}^{X}$ is given by
\begin{equation}
\tilde{I}_{(1)}^{X}=\sum_{i,j} M_{(1),ij}I^{X}_{1,1}.
\end{equation}

Now we have to perform the last coset extension from $\Gamma_{(123||456)}$ to
the full modular group
$\Gamma$. To do so we have to consider modular transformations
that take the spin structure $(123||456)$ into any other of the remaining
$9$ even spin structures. Because $\tilde{I}_{(1)}$ is the sum of 12 terms
and we have ten even spin structures, the modular invariant result will be
the sum of $120$ terms of the type $M_{(i),jl}I_{1,1}^{X}$
\begin{equation}
\sum_{e} I_{e} = \sum_{i=1}^{10}\sum_{j,k} M_{(i),jk} I^{X}_{1,1}.
\label{sum1}
\end{equation}
In fact it is somewhat convenient to reorder the previous expression as
a sum of $30$ terms of the form
\begin{equation}
(M_{(1),ij}+M_{(2),ij}+M_{(3),ij}+M_{(4),ij})I_{1,1}^{X},
\label{sum2}
\end{equation}
where $M_{(k),ij}$ ($k=2,3,4$) are obtained from $M_{(1),ij}$ by $3$ 
generators that leave invariant the pair $\{ij\}$. The point is that
all $M_{(k),ij}$ are proportional to ${\cal P}_{ij}/{{\cal P}_{12}}$.
From (\ref{Is}) and (\ref{tI}) we find that for example
\begin{equation}
\sum_{i=1}^{4} M_{(i),12}=1+\frac{\lambda_{1}^{3}\lambda_{2}^{3}(
\lambda_{3}-1)^{3}-\lambda_{3}^{3}(\lambda_{3}-1)^{3}-\lambda_{3}^{3}
(\lambda_{1}-1)^{3}(\lambda_{2}-1)^{3}}{(\lambda_{3}-\lambda_{1})^{3}
(\lambda_{3}-\lambda_{2})^{3}}.
\label{12}
\end{equation}
Using genus two Riemann theta
functions this reads
\begin{eqnarray}
\left\{\theta^{12}\left[
\begin{array}{cc}
0 & \frac{1}{2} \\
\frac{1}{2} & 0 
\end{array}
\right](0|\tau)\sum_{i=1}^{4} M_{(i),12}\right\}= \theta^{12}\left[
\begin{array}{cc}
0 & \frac{1}{2} \\
\frac{1}{2} & 0 
\end{array}
\right](0|\tau) 
- \theta^{12}\left[
\begin{array}{cc}
0 & \frac{1}{2} \\
0 & 0 
\end{array}
\right](0|\tau) \nonumber \\
+ \theta^{12}\left[
\begin{array}{cc}
\frac{1}{2} & 0 \\
0 & 0 
\end{array}
\right](0|\tau) - \theta^{12}\left[
\begin{array}{cc}
\frac{1}{2}& 0 \\
0 &\frac{1}{2} 
\end{array}
\right](0|\tau). \hspace*{1cm}
\end{eqnarray}

The complete expression for the genus two cosmological constant will
be the sum of $30$ terms of the type (\ref{12}) which can be obtained
from it by modular transformations. As a concrete example, using
the modular transformation $(12)$ we find
\begin{eqnarray}
\sum_{i=1}^{4} M_{(i),21} &=& -\left[1+\frac{\lambda_{1}^{3}\lambda_{2}^{3}(
\lambda_{3}-1)^{3}-\lambda_{3}^{3}(\lambda_{3}-1)^{3}-\lambda_{3}^{3}
(\lambda_{1}-1)^{3}(\lambda_{2}-1)^{3}}{(\lambda_{3}-\lambda_{1})^{3}
(\lambda_{3}-\lambda_{2})^{3}}\right] \nonumber \\
&\times& \frac{\lambda_{1}(\lambda_{3}-\lambda_{1})(\lambda_{2}-1)^{3}}{
\lambda_{2}(\lambda_{3}-\lambda_{2})(\lambda_{1}-1)}\frac{{\cal P}_{21}}{
{\cal P}_{12}}.
\label{piece2}
\end{eqnarray}

Then we have arrived at a modular invariant expression for the genus two
cosmological constant of the $24$ two-dimensional heterotic models under study.
As we expect from the fact that they are not supersymmetric, the integrand of
the cosmological constant does not vanish identically contrary to the case of
the ten-dimensional heterotic string \cite{tomas}. However the 
expression gotten (of which (\ref{12}) and 
(\ref{piece2}) are just a piece) is rather 
difficult to work with. To check whether or not $\Lambda_{2-loops}$ vanishes
we should integrate this expression to the fundamental domain in the 
$\lambda_{i}$-space which seems a rather scary and maybe impossible 
task. We will follow a different path and will turn to the computation of
the one-loop amplitude with two external massless states. This computation
hopefully will serve us in a double way; from it we can get the mass
corrections to the massless states in the theory and some indirect
information about the genus two cosmological constant could be extracted
along the lines of \cite{moore-bos}.

\section{The Two-Point Function for Massless Neutral \newline
Bosons at One Loop}
\label{nb}
\setcounter{equation}{0}

For the computation of the two-point function for two massless states 
we will use the bosonic construction and then our task will be four-fold,
since we will have to compute the amplitude for charged and neutral states
in the untwisted and twisted sectors of the theory. In this section we will
perform the computation for the states in the Cartan subalgebra of the
left-moving gauge group for both twisted and untwisted states leaving
for the next section the computation for charged states. 

In the B formulation of the model, the world-sheet action is
\begin{equation}
S[X^{\mu},X^{i},\psi^{\mu},\psi^{i},\phi^{I}]=
S_{2d}[X^{\mu},\psi^{\mu}]+S_{R,int}[X^{i},\psi^{i}]+
S_{L,int}[\phi^{I}],
\end{equation}
where 
\begin{eqnarray}
S_{2d}&=&-\frac{1}{8\pi}
\int d^{2}z [\partial_{z}X^{\mu}\partial_{\bar{z}}X_{\mu}
+2i\psi^{\mu}\partial_{\bar{z}}\psi_{\mu}], \nonumber \\
S_{R,int} &=& -\frac{1}{8\pi}
\int d^{2}z[\partial_{z}X^{i}\partial_{\bar{z}}X^{i}
+2i\psi^{i}\partial_{\bar{z}}\psi^{i}+\lambda_{R}\partial_{\bar{z}}
X^{i}\partial_{\bar{z}}X^{i}], \nonumber \\
S_{L,int} &=& -\frac{1}{8\pi}
\int d^{2}z[\partial_{z}\phi^{I}\partial_{\bar{z}}
\phi^{I}+\lambda_{L}\partial_{z}\phi^{I}\partial_{z}\phi^{I}]
\end{eqnarray}
and $\lambda_{L,R}$ are lagrange multipliers enforcing the chiral
character of the bosons. In what follows we will use units in which 
$\alpha^{'}=2$.

Let us begin with neutral untwisted states. The vertex operators in the
zero picture are
\begin{equation}
V_{0}^{I,i}(k;z)=\frac{\kappa}{\pi}J^{I}(\bar{z})
[\partial_{z} X^{i}+i(k_{\mu}
\psi^{\mu})\psi^{i}](z)
e^{ik_{\mu}X^{\mu}(z,\bar{z})}.
\end{equation}
Here $\mu=0,1$ is a space-time index and $i=1,\ldots,8$ labels the
eight internal dimensions in the right-moving sector. $\kappa$ is
the string coupling constant and the  
$J^{I}$ are any of the 24 
currents associated with the Cartan subalgebra of the left-handed
gauge group $G_{L}$ 
\begin{equation}
J^{I}(\bar{z})=i \partial_{\bar{z}}\phi^{I},
\end{equation}
where $\phi^{I}$ live in the 24-dimensional Niemeier lattice.

To compute the amplitude we have to evaluate the correlator of two vertex 
operators on the torus 
fixing simultaneously the spin structures $e$ for the world-sheet fermions 
and the boundary conditions $(\alpha^{m},\alpha^{n})$ 
for the bosons $X^{a}$, and then sum over $e$, $m$ and $n$. We have
\begin{eqnarray}
A^{IJ,ij}_{(m,n)}
(k)=\frac{\kappa^{2}}{\pi^{2}}
\sum_{e}C_{e}(m,n)\int_{\cal F}\frac{d^{2}\tau}{\tau_{2}} \int d^{2}z\,
\overline{\eta(\tau)}^{-24}\frac{\theta^{4} [e]}{4\eta^{12}}
\langle J^{I}(\bar{z})J^{J}(0)\rangle \nonumber \\
\times\langle[\partial_{z}X^{i}+
i(k\cdot\psi)\psi^{i}](z)[\partial_{z}X^{j}-i(k\cdot\psi)\psi^{j}](0)
\rangle_{e}^{(m,n)}
\langle e^{ik\cdot X(z,\bar{z})}e^{-ik\cdot X(0)}\rangle .
\label{amplit-gen}
\end{eqnarray}
The sub and superscripts in the second correlator indicate the
boundary conditions on the torus and all the correlators are computed
integrating over the matter fields in a fixed point of the moduli space. 
The first correlator can be computed by splitting $\phi^{I}$ into
a classical and a quantum piece $\phi^{I}=\phi_{cl}+\phi_{q}$ and then
summing over classical vacua and integrating the quantum fluctuations. 
Classical vacua are labeled by the vectors $P_{L}\in \Lambda_{24}$
\begin{equation}
\phi^{I}_{cl}(\bar{z})=\phi^{I}_{0}+2\pi P^{I}\int^{\bar{z}} \bar{\omega},
\label{decomposition}
\end{equation}
where $\bar{\omega}$ is the abelian 1-form on the torus. Using  
$\langle \partial_{\bar{z}}\phi^{I}_{q}\rangle=0$ we are left with
\begin{eqnarray}
\langle J^{I}(z)J^{J}(0)\rangle &=&
-\langle\partial_{\bar{z}}\phi^{I}_{q}(z)
\partial_{\bar{w}}\phi^{J}_{q}(0)\rangle \sum_{P\in \Lambda_{24}}
e^{-i\pi\bar{\tau} P^{2}_{L}} \nonumber \\
&-& (2\pi)^{2}
\sum_{P_{L}\in\Lambda_{24}}P_{L}^{I}
P_{L}^{J}e^{-i\pi\bar{\tau} P_{L}^{2}}.
\label{ff}
\end{eqnarray}
The quantum part is readily evaluated in terms of the
prime form $E(z,0)=\theta_{1}(z|\tau)/\theta_{1}^{'}(0|\tau)$
\begin{equation}
\langle\partial_{\bar{z}}\phi^{I}(\bar{z})\partial_{\bar{w}}
\phi^{J}(0)\rangle = -\delta^{IJ}\partial_{\bar{z}}\partial_{\bar{w}}
\ln{E(\bar{z},0)}=\delta^{IJ}\partial_{\bar{z}}^{2}\ln{E(\bar{z},0)}.
\end{equation}
At the same time, writing the second sum in (\ref{ff}) as the 
derivative of the theta function of $\Lambda_{24}$ we finally find 
\begin{equation}
\langle J^{I}(\bar{z})J^{J}(0)\rangle=
-\delta^{IJ}\partial_{\bar{z}}^{2}\ln{
\overline{E(z,0)}}\overline{
\Theta_{\Lambda_{24}}(0|\tau)}+
\frac{\pi}{6 i}\delta^{IJ}\frac{\partial}{\partial\bar{\tau}}
\overline{\Theta_{\Lambda_{24}}(0|\tau)}.
\end{equation}

The third correlator in (\ref{amplit-gen}) is easily seen to be equal to
\begin{equation}
\langle e^{ik\cdot X(z,\bar{z})} e^{-ik\cdot X(0)}\rangle=
e^{k^{2}\langle X(z,\bar{z})X(0)\rangle} = e^{k^{2}G(z,0)},
\end{equation}
where the boson propagator is
\begin{equation}
G(z,w)=-\ln{|E(z,w)|^{2}} + \frac{2\pi}{\tau_{2}}({\rm Im}z)^{2},
\end{equation}
and then
\begin{equation}
\langle e^{ik \cdot X(z,\bar{z})} e^{-i k\cdot X(0)}\rangle=
e^{\frac{2\pi k^{2}}{\tau_{2}}({\rm Im}z)^{2}} |E(z,0)|^{-k^{2}}.
\end{equation}

To finish, we are left with the computation of
\begin{equation}
\langle [ \partial_{z}X^{i}+i(k\cdot\psi)\psi^{i}](z)
[\partial_{w} X^{j}-i(k\cdot\psi)\psi^{j}](0)\rangle_{e}^{(m,n)},
\end{equation}
which reduces  to 
\begin{equation}
\langle\partial_{z}X^{i}(z)\partial_{w}X^{j}(0)\rangle^{(m,n)}
-k^{2}\delta^{ij}S_{e}(z,0)^{2}\Theta_{(m,n)},
\label{gg}
\end{equation}
where we have used the fermion propagator (Szeg\"o kernel)
\begin{equation}
-\langle \psi(z)\psi(w)\rangle=
S_{e}(z,w)=\frac{\theta[e](z-w|\tau)}{E(z,w)\theta[e](0|\tau)}.
\end{equation}
The first term in (\ref{gg}) has to be computed along the same lines as the 
$\langle JJ\rangle$ correlator, but now taking into account that
$X^{i}$ has boundary conditions twisted by $(\alpha^{m},\alpha^{n})$ along
the two homology cycles of the torus 
\begin{eqnarray} 
\langle\partial_{z}X^{i}(z)\partial_{w}X^{j}\rangle^{(m,n)}&=&
\delta^{ij}\partial^{2}_{z}\ln{E(z,0)}\Theta_{(m,n)}
(0|\tau) \nonumber \\
&+&\frac{\pi}{2i}\delta^{ij}\frac{\partial}{\partial \tau}
\Theta_{(m,n)}(0|\tau).
\end{eqnarray}

Putting  all the ingredients together we get the amplitude for fixed
boundary conditions $(m,n)$
\begin{eqnarray}
A^{IJ,ij}_{(m,n)}(k)&=&\frac{\kappa^{2}}{\pi^{2}}\delta^{IJ}\delta^{ij}
\sum_{e}C_{e}(m,n)\int_{\cal F} \frac{d^{2}\tau}{\tau_{2}}
\int d^{2}z\; \bar{\eta}^{-24}\frac{\theta^{4}[e]}{4\eta^{12}}
F_{L}(\bar{z}|\bar{\tau})\,e^{k^{2}G(z,0)} \times \nonumber \\
& &\left\{\partial^{2}_{z}\ln{E(z,0)} \Theta_{(m,n)}+
\frac{\pi}{2i}\frac{\partial}{\partial\tau}\Theta_{(m,n)}
-k^{2}S_{e}(z,0)^{2}\Theta_{(m,n)}\right\},
\label{cartan}
\end{eqnarray}
where $F_{L}(\bar{z}|\bar{\tau})$ is defined through
$\langle J^{I}(\bar{z})J^{J}(0)\rangle=\delta^{IJ} F_{L}(\bar{z}|\bar{\tau})$.

In all the calculation that have led us to (\ref{cartan}) we have 
maintained $k^{2}$ without implementing the on-shell condition for the 
massless bosons $k^{2}=0$. At face value if we set $k^{2}=0$ in 
(\ref{cartan}) we get rid of the term proportional $S_{e}^{2}$. However
one has to be very careful, since after performing the integral in 
$z$ we can have terms of the form $1/k^{2}$ which might cancel the
overall $k^{2}$ to give a finite result \cite{examples}. In the 
case of the $(0,0)$ sector this is not even needed, since we have
\begin{equation}
\sum_{e}C_{e}(0,0) \theta^{4}[e](0|\tau)S_{e}(z,0)^{2}=0,
\end{equation}
due to a Riemann identity \cite{Namazie-Narain-Sarmadi}. In the 
other sectors, however, we do not have any Riemann identity so 
we have to study the limit $k^{2}\rightarrow 0$. In principle the only
source of divergence in the integral over $z$ is the point $z=0$ 
in which the two insertion points collide. In fact it can be checked that
in the limit $k^{2}\rightarrow 0$ there is no cancellation of the 
prefactor $k^{2}$ and then we find that these part of the regularized integral 
vanishes in that limit. Taking $k^{2}=0$ in (\ref{cartan}) we have 
\begin{eqnarray}
A^{IJ,ij}_{(m,n)}(k^{2}=0)=\frac{\kappa^{2}}{\pi^{2}}\delta^{IJ}\delta^{ij}
\sum_{e}C_{e}(m,n)\int_{\cal F} \frac{d^{2}\tau}{\tau_{2}}
 \bar{\eta}^{-12}\frac{\theta^{4}[e]}{4\eta^{12}}\int d^{2}z\,
F_{L}(\bar{z}|\bar{\tau})\times \nonumber \\
\left\{
\partial^{2}_{z}\ln{E(z,0)}\Theta_{(m,n)}
(0|\tau) + \frac{\pi}{2i}\frac{\partial}{\partial \tau}
\Theta_{(m,n)}(0|\tau)\right\}. \hspace*{1cm}
\label{k2=0}
\end{eqnarray}

First of all, one has to check that this expression is modular 
invariant after summing over all boundary conditions $(m,n)$ and
spin structures. This is easily done taking into account that 
if $\Lambda$ is a $d$-dimensional self-dual lattice, then under 
$S:\tau \rightarrow -1/\tau$, together with the 
transformation of $\Theta_{\Lambda}(0|\tau)$ given in (\ref{tl}), we have
\begin{eqnarray}
\frac{\partial}{\partial \tau}\Theta_{\Lambda}\left[
\begin{array}{c}
a \\
b
\end{array}
\right](0|\tau) &\rightarrow &
\tau^{\frac{d}{2}+2} \frac{\partial}{\partial \tau}\Theta_{\Lambda}
\left[
\begin{array}{c}
-b \\
a
\end{array}
\right](0|\tau)+
\frac{d}{2}\tau^{\frac{d}{2}+1}\Theta_{\Lambda}\left[
\begin{array}{c}
-b \\
a
\end{array}
\right](0|\tau),  
\end{eqnarray}
and the prime form transforms according to  
\begin{equation}
\partial_{z}^{2}\ln{E(z,0)} \rightarrow \tau^{2}\partial_{z}^{2}
\ln{E(z,0)}+2\pi i\tau .
\end{equation}
The invariance of the amplitude 
under $T:\tau\rightarrow =\tau+1$ is also easy to show. 

The next step is obviously to compute the integral over $z$ in (\ref{k2=0}). 
The integral we are dealing with has the form
\begin{equation}
I^{(m,n)}=\int d^{2}z F_{L}(\bar{z}|\bar{\tau})F_{R}^{(m,n)}(z|\tau),
\label{intt}
\end{equation}
where $F_{R}^{(m,n)}(z|\tau)$ is the right-handed counterpart of the function
$F_{L}(\bar{z}|\bar{\tau})$ defined above. A useful thing to 
notice is that 
it is possible to rewrite $F_{L}$ as 
\begin{equation}
F_{L}=-
\overline{\Theta}_{\Lambda_{24}}\partial_{\bar{z}}\overline{\rho}
(z,\bar{z})+\overline{B}_{14}(\tau,\bar{\tau}),
\end{equation}
where $\rho(z,\bar{z})$ is given by (cf. \cite{Minahan,Clavelli})
\begin{equation}
\rho(z,\bar{z})=\partial_{z}\ln{E(z,0)}
+\frac{\pi}{\tau_{2}}(z-\bar{z})
\end{equation}
is a well defined function on the torus and 
\begin{equation}
B_{14}(\tau,\bar{\tau})=\frac{\pi}{\tau_{2}}\Theta_{
\Lambda_{24}}-\frac{\pi}{6i}\frac{\partial}{\partial\tau}
\Theta_{\Lambda_{24}}
\end{equation}
is independent of $z$ and transforms as a modular function of weight $14$.

In $F_{R}^{(m,n)}(z)$ we find the same structure than in $F_{L}$ and
therefore we can follow the same strategy and write
\begin{equation}
F_{R}^{(m,n)}= \Theta_{(m,n)}\partial_{z}\rho(z,\bar{z})
-B_{6}^{(m,n)}(\tau,\bar{\tau}),
\end{equation}
where now
\begin{equation}
B_{6}^{(m,n)}(\tau,\bar{\tau})=\frac{\pi}{\tau_{2}}\Theta_{(m,n)}-
\frac{\pi}{2 i}\frac{\partial}{\partial\tau}\Theta_{(m,n)}.
\end{equation}

Then $I^{(m,n)}$ reads
\begin{eqnarray}
I^{(m,n)}&=&-\overline{\Theta}_{\Lambda_{24}}\,
\Theta_{(m,n)}
\int d^{2}z \partial_{\bar{z}}\overline{\rho}
\partial_{z}\rho + \overline{B}_{14} 
\,\Theta_{(m,n)}\int d^{2}z
\partial_{z}\rho \nonumber \\
&+& B_{6}^{(m,n)}\, \overline{\Theta}_{\Lambda_{24}}
\int d^{2}z \partial_{\bar{z}}\overline{\rho}
-\overline{B}_{16}B^{(m,n)}_{6} \int d^{2}z.
\label{Imn}
\end{eqnarray}
Now all the integrals can be explicitely calculated at the price of 
losing holomorphic factorization. A special care
is needed in doing so, since the integrand in all the first three integrals
is singular at $z=0$ and the integrals are naively divergent. This divergence
corresponds to the point in which the insertions of the two vertex operators
come together. In order to regularize this divergence we are
going to cut off a small circle $|z|<\epsilon$ around $z=0$. Then we have
for the first integral
\begin{equation}
I_{1}=\int_{T_{\epsilon}^{2}} d^{2}z \partial_{\bar{z}}\overline{\rho}
\partial_{z}\rho = \frac{1}{2i}\int_{T_{\epsilon}^{2}}
\bar{\partial}\overline{\rho}\wedge
\partial \rho = \frac{1}{2i}\int_{|z|=\epsilon}dz\overline{\rho}\partial_{z}
\rho ,
\end{equation}
where we have used the crucial fact that $\partial\bar{\partial}\rho=0$.
For the crossed terms we have
\begin{equation}
I_{2}=\int_{T_{\epsilon}^{2}}d^{2}z\partial_{z}\rho=\frac{1}{2i}\int_{|z|=
\epsilon}d\bar{z}\,\rho .
\end{equation}
The last integral in (\ref{Imn}) is simply equal to the area of the torus,
minus the area of the removed circle namely 
$\tau_{2}-\pi\epsilon^{2}$. 

Since $I_{1}$ and $I_{2}$ are line integrals over
$|z|=\epsilon$ we only need to study the behavior of $\rho(z,\bar{z})$
near $z=0$. $E(z,0)$ can be written \cite{Lerche}
\begin{equation}
E(z,0)=z\exp{\left[-\sum_{k=1}^{\infty} \frac{z^{2k}}{2k}G_{2k}(\tau)\right]},
\end{equation}
with $G_{2k}(\tau)$ for $k>1$ the $k$-th Eisenstein series
\begin{equation}
G_{2k}(\tau)=\sum_{m,n\in{\bf Z}}{^{'}}(m\tau+n)^{-2k}
\end{equation}
and $G_{2}(\tau)$ the holomorphic-regularized Eisenstein series of
weight 2
\begin{equation}
G_{2}(\tau)=-\frac{1}{3}\frac{\theta_{1}^{'''}}{\theta_{1}^{'}}\, .
\end{equation}
Then we find the following Laurent series for
$\rho(z,\bar{z})$ 
\begin{eqnarray}
\rho(z,\bar{z})=\frac{1}{z}+\left[\frac{\pi}{\tau_{2}}(z-\bar{z})
-G_{2}\,z\right]-\sum_{k=2}^{\infty}z^{2k-1} G_{2k}.
\label{IL}
\end{eqnarray}
Substituting in $I_{1}$ and $I_{2}$ and performing the phase integral
we find
\begin{eqnarray}
I_{1}&=&-\frac{\pi}{\epsilon^{2}}+\frac{\pi^{2}}{\tau_{2}}+
\pi\left|\hat{G}_{2}\right|^{2}\epsilon^{2}
-\pi\sum_{k=2}^{\infty}\frac{\epsilon^{4k-2}}{2k}|G_{2k}|^{2}, \nonumber \\
I_{2}&=&\pi\epsilon^{2}\hat{G}_{2},
\end{eqnarray}
where now 
\begin{equation}
\hat{G}_{2}(\tau,\bar{\tau})=G_{2}(\tau)-\frac{\pi}{\tau_{2}},
\end{equation}
which is not holomorphic but transforms as a weight $2$ modular
function.
Mixing all the ingredients we finally arrive at
\begin{equation}
I^{(m,n)}=\frac{\pi}{\epsilon^{2}}\overline{\Theta}_{\Lambda_{24}}
\Theta_{(m,n)}-\frac{\pi^{2}}{\tau_{2}}\overline{\Theta}_{\Lambda_{24}}
\Theta_{(m,n)}-\tau_{2}\overline{B}_{14}B^{(m,n)}_{6}+O(\epsilon^{2}).
\label{ep}
\end{equation}

When computing the total amplitude we are going to have to sum over 
boundary conditions and spin structures, so we will need to evaluate 
the quantity $\sum_{e}\sum_{m,n}C_{e}(m,n)\theta^{4}[e]I^{(m,n)}$.
Using the definition of $B_{6}^{(m,n)}$ and the corresponding 
theta functions $\Theta_{(m,n)}$ as well as
some well-known results about the ring of modular
functions \cite{Koblitz} we find 
\begin{eqnarray}
\sum_{e}\sum_{m,n=0}^{1} C_{e}(m,n)\, \theta^{4}[e]\Theta_{(m,n)} 
&=& 96 \eta^{12},
\nonumber \\
\sum_{e}\sum_{m,n=0}^{1} C_{e}(m,n)\,\theta^{4}[e] B_{6}^{(m,n)} &=&
-96\eta^{12}\hat{G}_{2}, 
\end{eqnarray}
and 
\begin{eqnarray}
\overline{B}_{14} = -\overline{\hat{G}}_{2}
\overline{\Theta}_{\Lambda_{24}}+\frac{\pi^{2}}{6\zeta(14)}
\overline{G}_{14}. 
\end{eqnarray}
The final result is 
\begin{eqnarray}
& &\sum_{e}\sum_{m,n=0}^{1} C_{e}(m,n)\,\theta^{4}[e] I^{(m,n)} =
\frac{96\pi}{\epsilon^{2}}\,\eta^{12}\,\overline{\Theta}_{\Lambda_{24}} 
-\frac{96\pi^{2}}{\tau_{2}}
\eta^{12}\overline{\Theta}_{\Lambda_{24}} \nonumber \\
& &\hspace*{2cm}-96\tau_{2}\,\eta^{12}\,\left[
\overline{\Theta}_{\Lambda_{24}}
|\hat{G}_{2}|^{2}-\frac{\pi^{2}}{6\zeta(14)}\,\overline{
G}_{14}\hat{G}_{2}\right] +O(\epsilon^{2}).
\end{eqnarray}

Before going on any further, let us have a closer look at  our cutoff
$\epsilon$. We have regularized our integrals by removing a small 
circle with radius $\epsilon$ around $z=0$. Let us assume that 
we perform a modular transformation on our torus. In that case we
know that $z\rightarrow z/\tau$ and then we will have that after 
performing this transformation the boundary of our circle will
also shrink according to $\epsilon\rightarrow\epsilon/|\tau|$. So
in a sense we can say that $\epsilon$ is charged
under the modular group, since maintaining $\epsilon$ invariant under
a modular transformation would have the result of losing modular invariance
in the expansion in powers of $\epsilon$. It would be 
much more convenient to have a  neutral cutoff
under modular transformations. Let us look at the problem
in a more geometrical way; we want the radius $\epsilon$ of
the circle we remove from the torus to be small in order to use the series
expansion in powers of $z$ in the computation of the integrals. Nevertheless 
in the region in which $\tau_{2}\rightarrow 0$ we are dealing with very
small tori, and  $\epsilon$ must go to zero in order the circle to be a
well-defined neighborhood of $z=0$; if the circle is too large it 
will intersect with itself, since now the size of the torus shrink to 
zero. However we are not interested in having a scaling of $\epsilon$
just as $\sqrt{\tau_{2}}$, since in that case in the region in which
$\tau_{2}\rightarrow\infty$ (large tori) we would have that the area
of the circle would go to infinity although it can be small at the
scale of the torus. What we want is the radius $\epsilon$ to be arbitrarily
small, let us say of order $\tilde{\epsilon}\ll 1$, at all scales (i.e.,
all $\tau$) and to be at the same time small compared with the torus 
size which implies that $\epsilon$ must vanish when $\tau_{2}$ goes to
zero. These conditions can be accomplished if
we define our cutoff $\tilde{\epsilon}$ according to
\begin{equation}
\epsilon=\tilde{\epsilon}f(\tau,\bar{\tau}),
\label{cut}
\end{equation}
where $f(\tau,\bar{\tau})$ is of the order one at $\infty$, goes to
zero when $\tau_{2}\rightarrow\infty$, does not vanish anywhere else in
the upper half plane and it is such that under $S$ we have 
$f(\tau,\bar{\tau})\rightarrow f(\tau,\bar{\tau})/|\tau|$. $\tilde{
\epsilon}\ll 1$ is now our new neutral cutoff.
Let us notice that by expanding in powers of $\tilde{\epsilon}$ instead
of $\epsilon$ we do not modify the finite part but  we can 
write all the coefficients in the expansion as integrals over ${\cal F}$
of a modular invariant function. 

A first question that arises about $f(\tau,\bar{\tau})$ is whether or not
such a function exists. The easiest way to prove this 
existence theorem is just to construct a concrete example. Without much
effort one can find, for example,  
\begin{equation}
f(\tau,\bar{\tau})=2\left[\sum_{i=2}^{4}|\theta_{i}
(0|\tau)|^{2}\right]^{-1}.
\end{equation}
Indeed $f(\tau,\bar{\tau})$ transforms in the right way under the 
modular group and does not vanish
anywhere in the upper half plane, the theta series converging
for every $\tau$ such that $\tau_{2}>0$. Moreover, since the $\theta_{i}$'s
($i=2,3,4$) do not vanish in the upper half plane, 
$f(\tau,\bar{\tau})$
is finite in the same region. Of course it is quite easy to provide 
different examples for $f(\tau,\bar{\tau})$. We will further discuss 
this ambiguity in Sec. \ref{con}. 

Let us finally integrate over the fundamental region ${\cal F}$. The 
resulting $\tilde{\epsilon}$ expansion is
\begin{eqnarray}
A^{IJ,ij}(k^{2}=0)&=&\frac{24\kappa^{2}
F^{(-2)}[f]}{\pi\tilde{\epsilon}^{2}}
\delta^{IJ}\delta^{ij}+
\kappa^{2}\delta^{IJ}\delta^{ij}\Lambda_{1-loop}\nonumber \\
&-&
\frac{24\kappa^{2}}{\pi^{2}}F^{(0)}
\delta^{IJ}\delta^{ij}+O(\tilde{\epsilon}^{2}),
\label{Aab}
\end{eqnarray}
where $\Lambda_{1-loop}$ is the one-loop induced cosmological constant
with bosonic states contributing with a minus sign;
$F^{(-2)}[f]$ depends functionally on the  regulating function 
$f(\tau,\bar{\tau})$
\begin{equation}
F^{(-2)}[f]=\int_{\cal F}\frac{d^{2}\tau}{\tau_{2}}f(\tau,\bar{\tau})^{-2}
[\overline{j(\tau)}-720+r_{\Gamma}(1)]
\end{equation}
and $F^{(0)}$ is given by
\begin{eqnarray}
F^{(0)}=\int_{\cal F}d^{2}\tau\, \hat{G}_{2}\left\{\left[
\overline{j(\tau)}-720+r_{\Gamma}(1)\right]\overline{\hat{G}}_{2}
-\frac{\pi^{2}}{6\zeta(14)}\frac{
\overline{G}_{14}}{\overline{\eta}^{24}}\right\}.
\label{f}
\end{eqnarray}

Before closing this section we will compute the amplitude for two
external twisted states. The only change with respect to (\ref{cartan})
appears in $F_{R}^{m,n}$ since now the right-moving 
part of the vertex operator is
\begin{equation}
V_{0}^{P_{R}}=iP_{R}^{k}\psi^{k}e^{iP_{R}^{i}X^{i}(z)}.
\end{equation}
If the two external states
have internal momenta $P_{R}$ and $-P_{R}$ ($P_{R}^{2}=1$) we have
\begin{equation}
F_{R,tws}^{m,n}(z|\tau)=-S_{e}(z,0)E(z,0)^{-1}
\Theta_{(m,n)}(zP_{R}^{i}|\tau),
\end{equation}
the left moving part $F_{L}$ being just the one defined above. Summing
over $(m,n)$ and the spin structure and using some theta function gymnastic
one easily proves that
\begin{equation}
\sum_{e}\sum_{m,n=0}^{1} C_{e}(m,n)\theta^{4}[e] F_{R,tws}^{m,n}=
E(z,0)^{-2}\sum_{i=2}^{3}C_{i}\theta_{i}^{12}(0|\tau)
\frac{\theta^{2}_{i}(z|\tau)}{\theta_{i}^{2}(0|\tau)},
\label{t}
\end{equation}
$C_{2}=C_{4}=-C_{3}=1$.
This expression seems a little unpleasant. It is worth noticing, however,
that (\ref{t}) is a holomorphic doubly periodic function with a double
pole at $z=0$ and its Laurent expansion around this point has no term
in $z^{0}$. Using the results summarized in Appendix C we find
\begin{equation}
\sum_{e}\sum_{m,n=0}^{1} C_{e}(m,n)\theta^{4}[e] F_{R,tws}^{m,n}=
-96\eta^{12}{\cal P}\left(z\left|\frac{1}{2},\frac{\tau}{2}\right.\right),
\end{equation}
where ${\cal P}(z|\omega_{1},\omega_{2})$ is the Weierstrass elliptic
function with semiperiods $\omega_{1}$ and $\omega_{2}$. This can be
further related with the function $\rho(z,\bar{z})$ using the
identity
\begin{equation}
{\cal P}\left(z\left|\frac{1}{2},\frac{\tau}{2}\right.\right)=
-\partial_{z}\rho(z,\bar{z})-\hat{G}_{2},
\label{Wr}
\end{equation}
which allows us to write finally
\begin{equation}
\sum_{e}\sum_{m,n=0}^{1} C_{e}(m,n)\theta^{4}[e] F_{R,tws}^{m,n}=
96\eta^{12}\partial_{z}\rho(z,\bar{z})+96\eta^{12}\hat{G}_{2}.
\end{equation}
This is exactly the same result we got for the untwisted states
(notice the overall minus sign in the definition of $B_{6}$).

Our final result is that the two-point amplitude on the torus for two neutral
external states is given by (\ref{Aab}) and it has the same expression for 
twisted and untwisted external states. This is no wonder, since we know that
both kind of states in fact combine together in the adjoint representation
of $SU(2)^{8}$. It is also easy to check
that the result for the two-point
amplitude can also be obtained using the fermionic construction. In fact
it is clear that for example
(\ref{t}) can be rewritten in terms of fermion propagators
and interpreted as the correlation function of the vertex operators in 
the F construction.

\section{The Case of the Charged Bosons}
\label{nc}
\setcounter{equation}{0}

We now turn to the computation of the two point function 
for the $24\times r_{\Gamma}(1)$ charged states both twisted 
and untwisted. Since the calculation will be very similar to the one
made in the previous section we will skip here the details.
We can 
make use of the formula (\ref{amplit-gen}) but now we have to use a
different expression for the world-sheet currents $J(\bar{z})$. Charged
bosons are related with the simple roots of the corresponding gauge group. 
These roots are precisely the vectors $\alpha^{I}$ of the left-moving
lattice with $\alpha^{2}=2$. The current associated with the root
$\alpha$ is
\begin{equation}
J^{\alpha}(\bar{z})=c_{\alpha} e^{i\alpha\cdot\phi(\bar{z})},
\end{equation}
where $c_{\alpha}$ is a cocycle satisfying
\begin{equation}
c_{\alpha}c_{\alpha^{'}}=e(\alpha,\alpha^{'})\,c_{\alpha+\alpha^{'}} 
\end{equation}
and $e(\alpha,\alpha^{'})=0$ unless $\alpha+\alpha^{'}$ is a 
root. 

We are going to proceed as with the neutral bosons by writing 
 $\phi^{I}=\phi^{I}_{cl}+\phi^{I}_{q}$. Using (\ref{decomposition})
we get
\begin{equation}
\langle J^{\alpha}(\bar{z})J^{\beta}(0)\rangle=e(\alpha,\beta)c_{\alpha
+\beta}
\langle e^{i\alpha\cdot\phi_{q}(\bar{z})}e^{i\beta\cdot\phi_{q}(0)}\rangle
\sum_{P_{L}\in\Lambda_{24}} e^{-i\pi\bar{\tau}-2\pi i\, z\,\alpha\cdot P_{L}}.
\end{equation}
The integration over the zero mode $\phi^{I}_{0}$ gives raise to a
delta function that enforces $\alpha+\beta=0$ and that we will drop in the
following. We can write the contribution to the 
total amplitude in the sector $(m,n)$ as 
\begin{eqnarray}
A_{(m,n)}^{\alpha,ij}(k)=\frac{\kappa^{2}}{\pi^{2}}\delta^{ij}
\sum_{e}C_{e}(m,n)
\int_{\cal F}\frac{d^{2}\tau}{\tau_{2}}\int d^{2}z\, \bar{\eta}^{-24}
\frac{\theta^{4}[e]}{4\eta^{12}} F_{L}^{(\alpha)}(\bar{z}|\bar{\tau})
e^{k^{2}G(z,0)}\times & & \nonumber \\
\left\{\partial_{z}^{2}\ln{E(z,0)}\Theta_{(m,n)}
+\frac{\pi}{2i}\frac{\partial}{\partial\tau}\Theta_{(m,n)}
-k^{2}S_{e}(z,0)^{2}\right\}, & \hspace*{0.3cm} &
\end{eqnarray}
where 
\begin{equation}
F^{(\alpha)}_{L}(\bar{z}|\bar{\tau})=
\overline{\Theta_{\Lambda_{24}}(z\,\alpha^{I}|\tau)}\, \overline{
E(z,0)}^{\,-2}
\end{equation}
and we have applied $e(\alpha,-\alpha)=1$. 

The trick to deal with this integral is somewhat similar to the one
we used for the case of the twisted bosons. $F_{L}^{(
\alpha)}(z|\tau)$ is a holomorphic doubly periodic function on the
torus and then can be expressed in terms of the elliptic function,
which in turn we know how to write in terms of $\rho(z,\bar{z})$
(in this discussion we sill work with complex conjugate expressions 
in order to simplify the expression). Let us make
use of some general properties of $\Theta_{\Lambda_{24}}(z\alpha^{I}|
\tau)$. Any of the 23 (in this discusion  the Leech lattice is excluded)
Niemeier lattices is a Lie algebra lattice which, in general, is composed
of several factors ${\cal L}_{1}\times\ldots$ where ${\cal L}_{i}\neq
D_{1}$. If we take a base of orthonormal vectors we can label the
basis vectors in such a way that $\alpha^{I}$ lies in the 
$i$-th factor and  has coordinates $\alpha=(1,1,0,\ldots,0)$. Then it is 
easy to see that $\Theta_{\Lambda_{24}}(z\alpha^{I}|\tau)$ can be
written (we drop the arguments when theta functions are evaluated in $z=0$)
\begin{equation}
\Theta_{\Lambda_{24}}(z\alpha^{I}|\tau)=
C_{1}(\tau)\left[\frac{\theta_{1}(z|\tau)}{\theta_{1}^{'}(0|\tau)
}\right]^{2}
+\sum_{i=2}^{4} C_{i}(\tau)\left[\frac{\theta_{i}(z|\tau)}{\theta_{i}(0|\tau)
}\right]^{2}.
\end{equation}
The transformation properties of $C_{i}(\tau)$ can be gotten from the
ones for $\Theta_{\Lambda_{24}}(z\alpha^{I}|\tau)$, and by evaluating
the expression at $z=0$ we see that $\Theta_{\Lambda_{24}}=\sum_{i=2}^{4}
C_{i}(\tau)$. Multiplying by $E(z,0)^{-2}$ we find
\begin{equation}
\Theta_{\Lambda_{24}}(z\alpha^{I}|\tau) E(z,0)^{-2}=
C_{1}(\tau)+\sum_{i=2}^{4} C_{i}(\tau)\left[
\frac{\theta_{i}(z|\tau)}{\theta_{1}(z|\tau)}\frac{\theta_{1}^{'}(0|\tau)
}{\theta_{
i}(0|\tau)}\right]^{2}.
\end{equation}
We can now introduce the Weierstrass elliptic function by using
(\ref{apB})
\begin{equation}
\Theta_{\Lambda_{24}}(z\alpha^{I}|\tau) E(z,0)^{-2}=\Theta_{\Lambda_{24}}
{\cal P}\left(z\left|\frac{1}{2},\frac{\tau}{2}\right.\right)+
C_{1}(\tau)-\sum_{i=2}^{4} C_{i}(\tau)e_{i}.
\end{equation}
In fact it can be checked that the affine term in the last expression
is a modular form of weight $14$, and then
\begin{equation}
F_{L}^{(\alpha)}(z|\tau)=\Theta_{\Lambda_{24}}{\cal P}\left(z\left|
\frac{1}{2},\frac{\tau}{2}\right.\right)+\frac{\pi^{2}}{6\zeta(14)}G_{14},
\end{equation}
where the coefficient of $G_{14}$ is fixed by comparing the series 
expansions. Taking into account (\ref{Wr}) we can write
\begin{equation}
F_{L}^{(\alpha)}(\bar{z})=-\overline{\Theta}_{\Lambda_{24}}\partial_{\bar{z}}
\overline{\rho}-\overline{\Theta}_{\Lambda_{24}}\overline{\hat{G}}_{2}
+\frac{\pi^{2}}{6\zeta(14)}\overline{G}_{14}.
\end{equation}
With this expression for $F_{L}^{(\alpha)}$ and the results of the 
previous section we have
\begin{eqnarray}
& &\sum_{e}\sum_{m,n=0}^{1} C_{e}(m,n)\theta^{4}[e]F_{L}^{(\alpha)}
F^{(m,n)}_{R} = \frac{96\pi}{\epsilon^{2}}\eta^{12}\overline{\Theta}_{
\Lambda_{2}}-\frac{96\pi^{2}}{\tau_{2}}\eta^{12}\overline{\Theta}_{\Lambda_{
24}} \nonumber \\
& &-96\tau_{2}
\eta^{12}\left[\overline{\Theta}_{\Lambda_{24}}
|\hat{G}_{2}|^{2}-\frac{\pi^{2}}{6\zeta(14)}\eta^{12}
\overline{G}_{14}\hat{G}_{2}\right]+O(\epsilon^{2}).
\end{eqnarray}
Multiplying by all the prefactors in (\ref{amplit-gen}) and integrating
over the modular parameter we finally find
\begin{eqnarray}
A^{\alpha,ij}(k^{2}=0)&=& \frac{24\kappa^{2}
F^{(-2)}[f]}{\pi\tilde{\epsilon}^{2}}
\delta^{ij}+\kappa^{2}\Lambda_{1-loop}\delta^{ij}\nonumber \\
&-&\frac{24\kappa^{2}}{\pi^{2}}F^{(0)}\delta^{ij}+O(\tilde{\epsilon}^{2}).
\label{charged}
\end{eqnarray}

In the case of charged twisted states no
computation is necessary, since we have shown in Sec. \ref{nb} that the result
has to be the one for untwisted states. Then (\ref{charged})
is valid for twisted and untwisted charged states. 

\section{The Infrared Behavior and Mass Corrections}
\label{dis}
\setcounter{equation}{0}

In the preceding two sections we have computed the two-point function
on the torus for the states in the massless sector of the 24 two-dimensional
heterotic strings discussed in sec. \ref{2}. We have checked that
the one loop correlator $\langle V_{0}V_{0}\rangle$ gives the same result
for all the massless states (twisted or untwisted). This is not so surprising
if we take into account that twisted and untwisted states in the B
model add up to fill the adjoint representation of $SU(2)^{8}$ or that
all states are on the same footing in the F construction.

\begin{figure}
\let\picnaturalsize=N
\def\picsize{2in}
\def\picfilename{toro2.eps}
\ifx\nopictures Y\else{\ifx\epsfloaded Y\else\input epsf \fi
\let\epsfloaded=Y
\centerline{\ifx\picnaturalsize N\epsfxsize \picsize\fi
\epsfbox{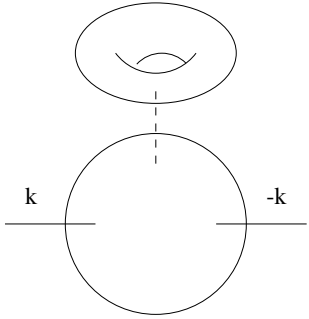}}}\fi
\caption{Factorization of the double-pole singularity}
\label{factor}
\end{figure}

However, in computing
the correlator of the two vertex operators on the torus we are
faced with the existence of divergences associated with the coincidence of
the two insertion points on the worldsheet (the $z\rightarrow 0$ limit). 
This divergence looks like the ones arising in the
computation of the one loop two-point graviton amplitude in the bosonic
string \cite{Minahan}. The standard interpretation in the literature 
of these kind of divergences is that they are
due to the propagation of an off-shell tachyon at zero momentum along the 
very long tube in fig. \ref{factor}, which shows the factorization
of the residue of the $1/\epsilon^{2}$ pole (see, for example, section 
8.2.4 in ref. \cite{GSW} and \cite{Minahan}).
Although this interpretation, in spite of involving off-shell guys, 
could be satisfactory for the bosonic string,
in our case it is very unpleasant to link the divergence with
the propagation of a tachyon since we are dealing with a  
tachyon free theory. Moreover, if we try to relate the residue   
of the $\epsilon^{-2}$ pole to the factorization limit shown in fig. 
\ref{factor} we find that the vertex operator in the $0$ picture must be
\begin{equation}
V^{tach}_{0}(k)\sim e^{i k\cdot X(\bar{z})},
\end{equation}
where $X^{\mu}(\bar{z})$ is the antiholomorphic part of the $X^{\mu}(
z,\bar{z})$ field. This vertex represents a rather weird state, being 
purely left-moving. Then the only conclusion to be extracted is that 
the interpretation of the divergence as caused by the propagation of 
off-shell tachyons is extremely unsatisfactory. 

Using the 
naive regularization of the amplitude in which a $\tau$-independent 
cutoff $\epsilon$ is introduced \cite{Minahan}, the final expansion
in powers of $\epsilon$ is not modular invariant, in the sense that 
the coefficients of $\epsilon^{n}$ cannot be written as integrals over
the fundamental domain of a modular invariant function, except for the
finite part with $n=0$.
This seems to be a problem, since modular invariance is
a necessary requisite for any sensible expression in string theory.
It is precisely this symmetry which allows us to interpret
any possible divergence appearing in any string amplitude as having an
infrared origin ($\tau_{2}\rightarrow \infty$) by excluding
the ultraviolet region. The breaking of 
modular invariance in the $\epsilon$-expansion then makes difficult to
see the divergence as due to an infrarred instability of
the theory. 

In our analysis we have shown that a modular invariant cutoff 
$\epsilon[f]=\tilde{\epsilon}f(\tau,\bar{\tau})$ can be introduced
 to provide  
a modular invariant expansion in powers of $\tilde{\epsilon}$.  
Now, however, the residue of the pole in 
$\tilde{\epsilon}^{-2}$ cannot be interpreted in terms of  
the propagation of an off-shell tachyon along the tube in fig. 
\ref{factor}, since this residue now depends functionally
on the {\it regulating function} $f(\tau,\bar{\tau})$. In a sense this
is satisfactory, since in a modular invariant description one expects
to project out any off-shell tachyons propagating in long tubes. The 
divergence
must then be interpreted in a different way; our theory, although tachyon 
free, is not finite and the arising divergence has an infrarred origin,
the only kind of divergences that any consistent string theory can contain.
The problem left is then to look for a way in which one can get rid of
this divergence. A first idea would be just to look for an analogue of the
Fischler-Susskind mechanism \cite{FS} which removes the logarithmic
divergences due to dilaton tadpoles by shifting the zero tree-level 
cosmological constant to the value induced at one loop. In our case
it is hard to find a similar mechanism since now there is no obvious
parameter in the sigma-model action whose analytic continuation could
absorb the one-loop divergences. In absence of a more elegant way to 
eliminate the divergence we will follow the procedure of ref. \cite{Minahan}
and just substract the pole. Then we find that, for both neutral and 
charged bosons, 
\begin{equation}
A_{2}=\kappa^{2}\Lambda_{1-loop}-\frac{24\kappa^{2}}{\pi^{2}}F^{(0)}.
\label{AA}
\end{equation}

It is worth stressing that the presence of this kind of divergences 
associated with the coincidence of the two insertions is ubiquous in
all the heterotic string models without space-time supersymmetry, since
the only way in which
one can get rid of them is when the integrand of the 
one-loop cosmological constant vanishes before integrating over the
fundamental region. This means that finitude seems to be a very 
difficult thing to get whenever we deal with 
non-supersymmetric heterotic string models.

Going back to $F^{0}$ as defined in ($\ref{f}$)
we can see that the term in the integrand proportional
to $\tau_{2}^{-2}$ is modular invariant by itself and proportional to the
one-loop cosmological constant. Then we can separate this term to get
\begin{eqnarray}
A_{2}&=& 2\kappa^{2}\Lambda_{1-loop}-\frac{24\kappa^{2}}{\pi^{2}}
\int_{\cal F}\frac{d^{2}\tau}{\tau_{2}}\left\{
-2\pi({\rm Re\,}G_{2})[\overline{j(\tau)}-720+r_{\Gamma}(1)] \right.
\nonumber \\
&+&\left. \frac{\pi^{3}}{6\zeta(14)}\frac{\overline{G}_{14}}{\overline{
\eta}^{24}}+
\tau_{2}
|G_{2}|^{2}[\overline{j(\tau)}-720+r_{\Gamma}(1)]-
\frac{\pi^{2}\tau_{2}}{6\zeta(14)}G_{2}
\frac{\overline{G}_{14}}{\overline{\eta}
^{24}}\right\}.
\label{shift}
\end{eqnarray}

Let us now turn to the modular integral in (\ref{shift}) and 
analyze the infrared region ($\tau_{2}\rightarrow\infty$).
We must remember that
in the neighborhood of $\tau=i\infty$
one must perform first the integral over
$\tau_{1}$, which enforces the level-matching condition, and then  
integrate $\tau_{2}$ all the way to infinity. Doing so we find that 
all unphysical tachyons cancel; however the integral is
infrared divergent. In fact we have a logarithmic 
divergence and a lineal one in the proper time when $\tau_{2}
\rightarrow \infty$.
These divergences are due to the fact that we are 
dealing with a two-dimensional system (cf. \cite{mpl2}). The term
$\Lambda_{1-loop}$ can be known exactly due to the remarkable properties
of the modular invariant function. Introducing   
an infrared cutoff in proper time $L^{2}$ to compute the integrals we find
\begin{eqnarray}
A_{2}&=&16\pi\kappa^{2}[12+r_{\Gamma}(1)]\ln{L^{2}}-
\frac{8\pi^{2}\kappa^{2}}{3}r_{\Gamma}(1)\,L^{2} \nonumber \\
&-&16\pi\kappa^{2}r_{\Gamma}(1)+\kappa^{2}A_{finite}+O(L^{-2}).
\label{analytic}
\end{eqnarray}
A numerical analysis of $A_{finite}$ yields
\begin{equation}
A_{finite}=-519.865+27.436\times r_{\Gamma}(1),
\label{numbers}
\end{equation}
where the numerical errors are in the third decimal place.

\begin{figure}
\let\picnaturalsize=N
\def\picsize{4in}
\def\picfilename{feynman.eps}
\ifx\nopictures Y\else{\ifx\epsfloaded Y\else\input epsf \fi
\let\epsfloaded=Y
\centerline{\ifx\picnaturalsize N\epsfxsize \picsize\fi
\epsfbox{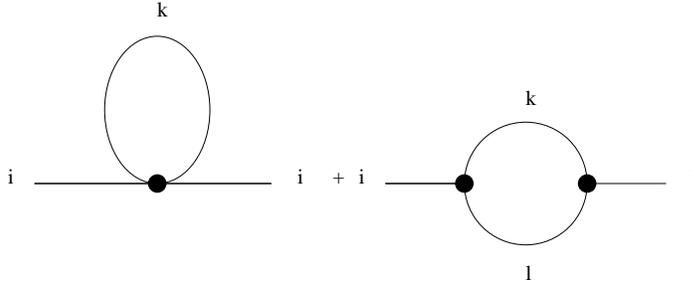}}}\fi
\caption{Feynman diagrams for the mass shift in the analog model}
\label{feynman}
\end{figure}

In fact, in order to understand 
this and the general structure of the two-point function, one can try to 
construct an analog model for (\ref{shift}). The term proportional to the
vacuum energy can be interpreted in a standard way as just the contribution
of the vacuum energies of the different fields with the subleties mentioned
in sec. \ref{2}; we will discuss this term later on when trying to compute
the mass corrections to massless states. Then let us center ourselves in the 
truncated amplitude $\tilde{A}$ without this term. Let us go to the
region of large $\tau_{2}$ which corresponds to very long tori.
In such a situation we can consider that only on-shell string states
circulate in the loop since in that region we can impose the
left-right level matching condition by integrating over $\tau_{1}$. Then
we can write (trading $\tau$ by $\tau_{1}+is$)
\begin{eqnarray}
\tilde{A} &=& -\frac{24\kappa^{2}}{\pi}\int_{\mu^{-2}}^{\infty}
\frac{ds}{s}\int_{-\frac{1}{2}}^{\frac{1}{2}} d\tau_{1}\left\{
2({\rm Re\,}G_{2})[\overline{j}-720+r_{\Gamma}(1)]-\frac{\pi^{2}}{6\zeta(14)}
\frac{\overline{G}_{14}}{\overline{\eta}^{24}}\right\}
\nonumber \\
&+&\frac{24\kappa^{2}}{\pi^{2}}\int_{\mu^{-2}}^{\infty}ds
\int_{-\frac{1}{2}}^{\frac{1}{2}}d\tau_{1}\left\{|G_{2}|^{2}[\overline{j}-720
+r_{\Gamma}(1)]-\frac{\pi^{2}}{6\zeta(14)}G_{2}\frac{\overline{G}_{14}}{
\overline{\eta}^{24}} \right\}.
\label{s}
\end{eqnarray}
Each integrand is a power series of the type $\sum_{m,n} a_{mn}e^{2\pi 
im\tau_{1}}e^{-2\pi n s}$ and the integration over $\tau_{1}$ restricts this 
sum to the $m=0$ terms in such a way that we can write
\begin{equation}
\tilde{A}=\sum_{k}\tilde{V}_{i,i,k,k}\int_{\mu^{-2}}^{\infty}
\frac{ds}{s} e^{-m_{k}^{2} s}+\sum_{k} \tilde{V}_{i,k,l}
\tilde{V}_{k,l,i}
\int_{\mu^{-2}}^{\infty} ds \,e^{-m^{2}_{k} s}.
\end{equation}
Here $V_{i,i,k,k}$ and $V_{i,k,l}$ are effective couplings which in
principle could be read from (\ref{s}) and the  
sum is over all the states running in the loop.
In fact, such a general structure for $\tilde{A}$ can be
obtained from the Feynman diagrams in fig. \ref{feynman}.
Now we can make an effective field theory interpretation of our result. In the
large proper time limit the truncated two point function $\tilde{A}$
is the sum of two
contributions. One of them comes from the degeneration of the torus into
a four point function on the sphere with two of the states joined by a 
long tube (first diagram in fig. \ref{feynman}). The second one has its
origin in a degeneration of the torus in which we have two three-point
functions on the sphere joined by two long tubes. As a matter of fact
the effective coupling $V_{i,j,k,l}$  must only include the 
$\alpha^{'}$ leading
contribution to the four-point tree level function; 
this diagram gives
\begin{equation}
\frac{V_{i,i,k,k}}{4\pi}\int_{0}^{\infty} \frac{ds}{s} e^{-m^{2}_{k}s},
\end{equation}
with $m$ the mass of the state running in the loop. The contribution 
coming from the second diagram depends on whether or not the masses of
the two internal states are equal. If they are not we have
\begin{equation}
\frac{V_{i,k,l}V_{k,l,i}}{4\pi(m_{l}^{2}-m_{k}^{2})}
\int_{0}^{\infty} \frac{ds}{s}
e^{-m_{l}^{2}s} + \frac{V_{i,k,l}V_{k,l,i}}{4\pi(m_{k}^{2}-m_{l}^{2})}
\int_{0}^{\infty} \frac{ds}{s}
e^{-m_{k}^{2}s}.
\end{equation}
But  when the masses of the two states coincide ($m_{k}=m_{l}$) we have
\begin{equation}
\frac{V_{i,k,l}V_{k,l,i}}{4\pi}\int_{0}^{\infty} ds\, e^{-m_{k}^{2}s}.
\end{equation}
From this and the Feynman diagrams in fig.
\ref{feynman} it is quite obvious why the effective vertex 
$V_{i,j,k,l}$ does not include subleading corrections in
$\alpha^{'}$. These corrections are included in the contribution
of the second diagram when $m_{k}$ and $m_{l}$ are not equal.

Now the origin of the divergences in (\ref{shift}) is clear. Due to
the low value of the dimension we will have divergences associated
with large $s$ which are logarithmic for the first diagram and logarithmic 
and 
linear for the second one; the divergent parts as $s\rightarrow 
\infty$ are
\begin{equation}
16\pi \kappa^{2}[12+r_{\Gamma}(1)] \ln s 
-\frac{8\pi^{2}\kappa^{2}}{3}r_{\Gamma}(1)\,s.
\label{ref}
\end{equation}
This kind of infrared divergences appear in massless field theory whenever 
$d\leq 2$ since in that case the measure in the Feynman integrals cannot cancel
the divergence in the propagator as the internal momentum in the loop 
goes to zero. 

In fact we can connect the general structure of the singular part
with what we know from the low energy field theory. As an example
let us consider that we have neutral external particles. In this case
according to the computations of sec. \ref{field} the leading contribution
in $\alpha^{'}$ comes from both diagrams with a charged particle
(with left and right-moving charge) running in the loop. 
Since the number
of such states is proportional to $r_{\Gamma}(1)$ we expect to have both
a linearly and a logarithmic 
divergent term proportional to $r_{\Gamma}(1)$ as we indeed have in 
(\ref{ref}). In addition, we also have a contribution coming from
the second diagram with one massive and one massless particles running 
in the loop, both charged only  with respect to $SU(2)^{8}$, which corresponds
to $O(\alpha^{'})$ terms in (\ref{4}). This gives a contribution
to the logarithmic singularity which is independent of $r_{\Gamma}(1)$.
Of course, in order to reproduce the concrete numbers in (\ref{ref})
one should sum over all the subleading constributions. In any case we see that
the structure of the result agrees qualitatively
with the analysis done in previous
sections. At any rate 
it must be clear that this field theoretical interpretation of 
the stringy result is by no means complete in the sense
that it cannot reproduce the exact result (\ref{shift}). In fact the
lesson we learned from the study of the partition function is that
no field theoretical description of a string amplitude can reproduce
the string theory calculation unless {\it intruder} (i.e., ghost-like) 
states are introduced in the game \cite{MM1,MM2}. The analysis of the 
previous paragraph 
is simply intended to give a more physical insight of the stringy result 
in terms of quantum fields.

One of the most interesting informations that can be extracted from
our computation of the genus one two-point function is the existence
of mass renormalization of the massless states. The point, however, is
a little bit subtle for  the 23 models without Atkin-Lehner symmetry. 
The interpretation of the on-shell two point function as quantum corrections
to the tree-level mass for the massless states of the string can only
be direct in the case of vanishing cosmological constant, since
only in this case the perturbative expansion is consistent in the sense
that the tree-level vacuum is also a good vacuum at one-loop level. 
In the case of models with one-loop induced vacuum energy ($r_{\Gamma}(1)
\neq 0$ in our case) the tree level vacuum is flat,
but after the inclusion of the one-loop effects this vacuum
no longer satisfies the equations of motion of the string, since now the
string is propagating in a (Anti-)de-Sitter space-time. It is for this reason
that only when $\Lambda_{1-loop}=0$ we can write 
\cite{Weinberg,Minahan2,Minahan}
\begin{equation}
\delta m^{2}_{i}= \left.\frac{ }{ }-\langle 
V_{0}^{i}(k)V_{0}^{i}(-k)\rangle
\right|_{k^{2}=0}-{\rm massless \,\, tadpoles}.
\end{equation}
In the case of the model with Atkin-Lehner symmetry it is easy to see that
all possible massless tadpoles vanish and then 
the mass shift for the states in the massless sector
of this  model is ($\alpha^{'}=2$)
\begin{equation}
\delta m^{2}_{Leech}\approx 519.865\kappa^{2}.
\end{equation}
Then we see that the massless sector does not survive the quantum corrections
in the string coupling constant.
For the remaining 23 models, things are not so easy as we have explained.
In fact, when quantizing a scalar 
field $\phi$
theory in curved space-time one must allow for a term in the action of the 
form $\xi\phi^{2}(x)R(x)$ where $R(x)$ is the scalar curvature and 
$\xi$ is a coupling constant \cite{Birrel-Davies}. 
If we have our field propagating in a (Anti-)De-Sitter space-time with
constant curvature $R$, the two point function contributes
to the renormalization of the wave function, the mass and $\xi$ 
according to 
\begin{equation}
A_{2}^{1-loop}(p^{2})=\delta Z\,p^{2}-\delta m^{2}-\delta\xi\,R.
\end{equation}
In our case this translates into a one-loop induced term in the 
effective action
of the form
\begin{equation}
S_{1-loop}=\int d^{2}x \left[\frac{1}{2}\delta m^{2} {\rm Tr\,}\Phi^{2}
+\frac{1}{2}\delta\xi\,R\,{\rm Tr\,}\Phi^{2}\right]\,.
\end{equation} 
Now we have to relate the scalar curvature with the parameters in our
models. At tree-level we are perturbing around a vacuum in which
all low-energy fields and the cosmological constant vanish (flat space-time). 
At one loop, however, our new vacuum
has $\Lambda\neq 0$ but none of the $\Phi$ fields gets a vacuum 
expectation value, so from the
dilaton beta function we must have $R\sim 2\Lambda$. In fact since the one
loop cosmological constant is proportional to $r_{\Gamma}(1)$ we have
that $R\sim r_{\Gamma}(1)$. This means that all the finite 
terms in (\ref{analytic})
which are proportional to $r_{\Gamma}(1)$ may be readsorbed in a
renormalization of $\xi$. In this way we find for the 23 models with
non-vanishing cosmological constant (here we do not have massless tadpoles
either)
\begin{eqnarray}
\delta m^{2} &\approx & 519.865 \kappa^{2}, \nonumber \\
\delta \xi &\approx & 0.454 \kappa^{2}.
\end{eqnarray}
So the mass renormalization for the massless states would be
the same for all the 24 two-dimensional heterotic models. 

One can wonder about the possibility of
having any breakdown of gauge symmetry because of these non vanishing
mass corrections.
To clarify this point the best thing to do is to go to the analogous
situation in field theory, that is, a theory with $N$ scalar fields
in the adjoint representation of the gauge group. This theory can be
viewed as the result of dimensional reduction of Yang-Mills theory in
$d+N$ to $d$ dimensions where the scalars appear as the $N$ internal
components of the gauge bosons. It is easy to see that the masses of
such scalars are not protected by any Ward identity, since after 
dimensional reduction the gauge parameter loses any dependence in the 
internal coordinates and then the internal components of the gauge field
are invariant under gauge transformations. A different problem is
how non-vanishing and infrared divergent two-point functions for 
propagating gauge fields affect gauge invariance in non-supersymmetric
string theories such as $SO(16)\times SO(16)$. This can only be addressed
by studying the string Ward identities for such amplitudes and the
possible anomalies that could arise in regularizing the amplitudes 
\cite{maharana}.

\section{Conclusions}
\label{con}
\setcounter{equation}{0}

We have tried to clarify how quantum corrections
in the string coupling constant modify the tree level structure of 
two-dimensional heterotic strings without space-time supersymmetry.
We have found that the 24 models constructed from the left-moving
bosonic string compactified on a Niemeier 
lattice and the right moving heterotic string
on $\Gamma_{8}$ modded out by the operator $\alpha$ defined in Sec. \ref{2}
they all have a right-moving level 2, $SU(2)^{8}$  gauge symmetry. 
Using
this fact we have been able to relate this bosonic construction of 
the right moving sector with a new one in terms of 
free worldsheet fermions. In ref. \cite{ZfP} a theorem was proved 
stating that for any two-dimensional heterotic string the partition function
has to be of the form
\begin{equation}
Z=Z_{R}\int_{\cal F}\frac{d^{2}\tau}{\tau_{2}^{2}}[\overline{j(\tau)}
-720+r_{\Gamma}(1)],
\end{equation}
where $Z_{R}$ must be a constant. This means that any two-dimensional
heterotic string either is supersymmetric or at most supersymmetry is 
broken only at the massless level. We will see how this result constraint
the possible fermionic constructions.

Let us consider the right-moving sector of the two-dimensional heterotic
string as formed by a set of 24 free world-sheet Majorana-Weyl fermions.
In Sec. \ref{field} we said how such theories are classified by 
a pair of semi-simple Lie groups $G$ and $H$, $H\subset G$
\cite{Antoniadis}.
The fact that our models live in two dimensions
forces ${\rm dim}\,G=24$ and the dimension of $H$ will 
determine the mass of the lowest lying fermion in the model. Since
from modular invariance we know that
supersymmetry can at most be broken only for the massless states we
can only have ${\rm dim\,}H=8,24$. In the first case we single out one
group of 8 world-sheet fermions transforming in the adjoint of $H$ and
project down to $(-1)^{F_{pseudo}}=1$. It is not difficult to realize 
that the lowest lying 
Ramond state is massless and we have the supersymmetric model ($Z_{R}=0$).

If ${\rm dim\,}H=24$ we take the same GSO projection over all 
the fermions and it can be easily seen that all massive levels
have the same number of fermionic and bosonic degrees of freedom except 
for the massless sector in which there are only 24 bosons in the adjoint
representation of $H$ ($Z_{R}=24$).  
Then we have seen that the only freedom we are left when constructing
two-dimensional heterotic models is (besides the choice of the
24-dimensional lattice) the election of the right-moving gauge group.
Different choices will differ in the actual couplings between
the low-energy fields although (\ref{action}) retains its general form.
However other aspects such as the one-loop two point function or
to some extend the two-loops cosmological constant appears to be
quite independent of the model chosen. 
In the present paper we have centered ourselves in the study of one of
these possibilities, namely the case $G=H=SU(2)^{8}$ since this is the 
model that results from the usual constructions in 
the previous literature \cite{Moore,MM1}.

We have studied the genus two cosmological constant for the $SU(2)^{8}$ model
and found, using the technique developed in \cite{tomas}, a modular invariant
expression that does not vanish before integration on the fundamental 
region of $Sp(2,{\bf Z})$. The question of the vanishing of this expression
after integration over the harmonic ratios $\lambda_{i}$ for the 
model with Atkin-Lehner symmetry ($r_{\Gamma}(1)=0$) seems difficult
to answer due to the unmanageable form of the integrand. However 
it is possible to give some indirect evidences that in fact this 
is not to be expected. The contribution to the genus two cosmological 
constant of Riemann surfaces in which a non-trivial homology
cycle is pinched off could be written as \cite{moore-bos}
\begin{equation}
\Lambda_{2-loop} \sim \sum_{i} d_{i}\int_{\mu}^{\infty}
\frac{ds}{s}\,e^{-m^{2}_{i}s}A_{1-loop}^{i,i},
\label{dg}
\end{equation}
where the sum is over all states in the string, $A_{1-loop}^{i,i}$ 
is 
the one-loop two-point function for the $i$-th state with mass $m_{i}$
and $d_{i}$ is a degeneration factor that takes into account the 
number of physical
degrees of freedom for each state. The boundary of the moduli space of
genus two Riemann surfaces has two branches. One of them (${\cal B}_{1}$) is 
parametrized
by the period matrix $\tau_{ij}$ when one of its diagonal entries 
goes to $i\infty$ (for example $\tau_{11}$). Geometrically 
this corresponds to the degeneration of a non-trivial homology  cycle.
The second branch (${\cal B}_{2}$) contains Riemann surfaces for which 
$\tau_{12}\rightarrow 0$, i.e., the trivial homology
cycle is degenerated. Over ${\cal B}_{1}$ the
genus two partition function takes the form (\ref{dg}) where $s \sim
\tau_{11}$, $\tau_{12}$ is the relative coordinate of the two-insertions
and $\tau_{22}$ is the modular parameter of the remaining torus 
\cite{moore-bos}. From our study of the genus one two-loop point function
for massless states we know that they are divergent not only in the limit 
of coincidence of the two insertions $\tilde{\epsilon}=0$ but also 
when $\tilde{\epsilon}\neq 0$ because of the low number of open 
space-time dimensions. Then we see that $\Lambda_{2-loop}$ will have a 
divergent contribution not only from ${\cal B}_{1}\cap {\cal B}_{2}$ 
but also from ${\cal B}_{1}$. $\tilde{\epsilon}$ can be seen as a 
{\it coordinate} over ${\cal B}_{1}$ in a neighborhood of 
${\cal B}_{1}\cap {\cal B}_{2}$. This would suggest that the integrated 
genus two cosmological constant is divergent due to the same kind
of infrared divergences that appear in the computation of one-loop 
scattering amplitudes.

In the study of the one-loop two point functions for these models we 
have found that in general they do not vanish. In fact during the
computation we have been faced with divergences associated with 
the coincidence of the two insertions. We have studied the origin of
this divergence and argued that it cannot be explained in terms of 
off-shell tachyons propagating along degenerated Riemann surfaces.
Using a modular invariant regulator we have identified the origin of
this divergence as an infrared instability of the theory. Even after 
subtracting the pole $\tilde{\epsilon}^{-2}$ we have found that 
the {\it finite} part of the regulated amplitude
is further afflicted from infrared divergences due to the fact
that the string lives in a two-dimensional target space-time. Moreover, since 
these infinities are caused by massless states living 
in two dimensions, they are also present in the low-energy
field theory described by the action (\ref{action}). This is contrary to
the infrared divergence associated with the pole  $\tilde{\epsilon}^{-2}$. 
In this latter case, the divergence is due to the 
coincidence of two composite operators in the two-dimensional field
theory on the world-sheet and then from a world-sheet point of view is of 
ultraviolet origin. However, looking at it from the two-dimensional 
target space
the divergence is infrared and involves the full string theory. Then 
it cannot have any counterpart in the low-energy effective theory for 
the massless modes, since here we are integrating out all the massive states; 
it is a purely stringy infrared divergence.  

In the case of the model based on the Leech lattice we have
computed the one loop mass-shift and found it to be positive. For the
other 23 models the interpretation of the one-loop two-point function
as the first quantum correction to the mass of the state is rather
problematic, since for them there is a one-loop induced cosmological
constant and then the one loop vacuum  does not satisfy the 
string equations of motion \cite{Minahan,Minahan2}. We argue that
the two-point
function then contributes not only to the mass renormalization but also to
the renormalization of the coupling $\xi$ between the massless scalar
fields and the scalar curvature in the low-enegy field theory. Identifying
the terms in the two-point function proportional to $R\sim r_{\Gamma}(1)$ with 
the renormalization of $\xi$ we find that the renormalization of the 
mass for the massless states would be the same for all the 24 models. 

\section*{Acknowledgements}

One of us (M.A.V.-M.) whises to thank J.\,L.\,F. Barb\'on and E. Witten 
for useful discussions and J. Maharana for conversations about his work 
on string Ward identities \cite{maharana}. The work of (M.A.V.-M.) has 
been supported by a Spanish MEC postdoctoral fellowship.

\section*{Appendix A: Theta Functions for Lattices}
\appendix

\def\theequation{A.\arabic{equation}}
\setcounter{equation}{0}

In this Appendix we will summarize some results about theta functions for
lattices \cite{DP,Serre}.
The theta series associated with a lattice $\Lambda$ is 
defined to be \cite{Serre}
\begin{equation}
\Theta_{\Lambda}(0|\tau)=\sum_{P\in\Lambda}
e^{i\pi\tau P\cdot P}.
\end{equation}
This definition can be easily generalized to include non-vanishing
first argument and characteristics. Let be $a^{I}$
and $b^{I}$ two vectors not in $\Lambda$ but such that
$2a,2b\in\Lambda$. Then we define
\begin{equation}
\Theta_{\Lambda}\left[
\begin{array}{c}
a \\ 
b
\end{array}
\right](v^{I}|\tau)=\sum_{P\in\Lambda} e^{i\pi\tau (P+a)\cdot (P+a)+
2\pi i\,(P+a)\cdot(v+b)}.
\end{equation}
In the case of null characteristics we will simply denote the corresponding
theta function by $\Theta_{\Lambda}(v^{I}|\tau)$.
Notice that since the sum is extended to all vectors in $\Lambda$
the characteristic vectors $a^{I}$ and $b^{I}$ are defined modulo shifts
by vectors in the lattice, i.e., they can be taken to live in 
$(\Lambda/2)/\Lambda$.

Let us consider from now on that $\Lambda$ is an even, self-dual lattice. 
Then under
the modular group generators $T$ and $S$ the theta functions behave as
\begin{eqnarray}
\Theta_{\Lambda}\left[
\begin{array}{c}
a \\
b
\end{array}
\right](v^{I}|\tau+1) &=& e^{-i\pi a^{2}}\Theta_{\Lambda}\left[
\begin{array}{c}
a \\
a+b
\end{array}
\right](v^{I}|\tau),
\nonumber \\
\Theta_{\Lambda}\left[
\begin{array}{c}
a \\
b
\end{array}
\right]\left(\left.\frac{v^{I}}{\tau}\right|-\frac{1}{\tau}\right)&=&
(-i\tau)^{\frac{d}{2}} e^{i\pi\frac{v^{2}}{\tau}} \Theta_{\Lambda}
\left[
\begin{array}{c}
-b \\
a
\end{array}
\right](v^{I}|\tau).
\label{tl}
\end{eqnarray}
It is also useful to derive the quasiperiodicity properties; given
$p^{I},q^{I}\in\Lambda$
\begin{equation}
\Theta_{\Lambda}\left[
\begin{array}{c}
a \\
b
\end{array}
\right](v^{I}+\tau p^{I}+q^{I}|\tau)=
e^{-i\pi\tau q^{2}-2\pi i\,p\cdot(v+b)+2\pi i a\cdot q}
\Theta_{\Lambda}\left[
\begin{array}{c}
a \\
b
\end{array}
\right]
(v^{I}|\tau).
\end{equation}

Lattice theta functions with characteristics $\{a^{I},b^{I}\}$ will be
even as  functions of $v^{I}$ if and only if $4 a\cdot b$ is an even
integer. If this is not the case then the theta function will be odd and
in particular will vanish at $v^{I}=0$. The zeroes of the 
$\Theta_{\Lambda}(v^{I}|\tau)$ are actually related to the existence of odd
characteristics; given
two vectors $\delta_{1}^{I},\delta_{2}^{J}
$ such that $2\delta_{1},
2\delta_{2}\in\Lambda$ and $4\delta_{1}\cdot\delta_{2}\in 2{\bf Z}+1$, we have
\begin{equation}
\Theta_{\Lambda}(\tau\delta_{1}^{I}+\delta_{2}^{I}|\tau)=0.
\end{equation}
This formula can be checked by writing the theta function with 
characteristics $\{\delta_{1}^{I},\delta_{2}^{I}\}$ in terms of 
$\Theta_{\Lambda}(v^{I}+\tau\delta_{1}^{I}+\delta^{I}_{2}|\tau)$ and
taking into account that, being an odd function of $v^{I}$, 
it has to vanish at $v^{I}=0$.

\section*{Appendix B: Riemann Surfaces in Hyperelliptic \newline
 Formalism and
the Knizhnik Formula}
\appendix

\def\theequation{B.\arabic{equation}}
\setcounter{equation}{0}

A genus $g$
hyperelliptic surface is defined as a two-dimensional surface that
uniformizes \cite{farkas-kra}
\begin{equation}
y(z)^{2}=\prod_{i=1}^{2g+2}(z-a_{i}),
\end{equation}
where $a_{i}=z(P_{i})$ with $z$ are holomorphic coordinates in ${\bf C}P^{1}$.
Every Riemann surface with $g\leq 2$ is hyperelliptic. Using a 
$SL(2,{\bf C})$ transformation we can fix the locations of three branching
points, the canonical choice being $a_{2g}=0$, $a_{2g+1}=1$ and $a_{2g+2}=
\infty$. The remaining $2g-1$ points on ${\bf CP}^{1}$ provide us with
good coordinates in the moduli space ${\cal M}_{g}$ of genus $g$ hyperelliptic
Riemann surfaces and then we have ${\rm dim\,}_{\bf C}\,{\cal M}_{g}=2g-1$.

\begin{figure}
\let\picturenaturalsize=N
\def\picsize{0.5in}
\def\picfilename{bitorus.eps}
\ifx\nopictures Y\else{\ifx\epsfloaded Y\else\input epsf \fi
\let\epsfloaded=Y
\centerline{\ifx\picnaturalsize N\epsfxsize \picsize\fi
\epsfbox{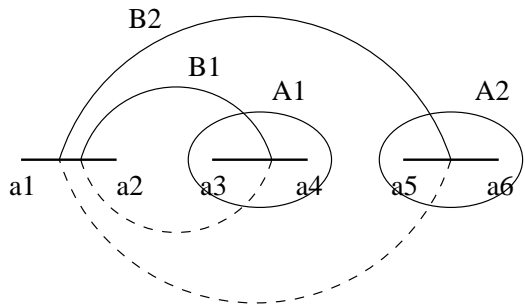}}}\fi
\caption{Genus-two Riemann surface.}
\label{bitorus}
\end{figure}

Let us focus on the $g=2$ case. We have 6 branch points and
the complex dimension of the moduli space is 3  which is the number of complex 
parameters of the genus-two period matrix. We represent this surface
schematically in fig. \ref{bitorus} with the basis for the homology 
cycles. Modular transformations in the hyperelliptic language amounts
to permutations of the branching points $a_{i}$ and then the five generators 
of the genus-two modular group are in one-to-one correspondence with the
generators of the braid group on the sphere $B_{5}$.
On the other hand the ten even spin structures on a genus two Riemann
surface are in one-to-one correspondence with partitions of the set of the
six branch points $(P_{1},\ldots,P_{6})$ into two subsets of three points
$(P_{i_{1}},P_{i_{2}},P_{i_{3}}||P_{j_{1}},P_{j_{2}},P_{j_{3}})$.
The exact correspondence can be found in the Appendix A of \cite{tomas}.
For $e=(P_{i_{i}},P_{i_{2}},P_{i_{3}}||P_{j_{1}},P_{j_{2}},P_{j_{3}})$
the corresponding theta function can be written in terms of the $a_{i}$'s
using Thomae's formula
\begin{equation}
\theta^{8}[e]=(\det{\sigma})^{-4}\prod_{k,l}^{g+1} a_{i_{k}i_{l}}^{2}
a_{j_{k}j_{l}}^{2}\,,
\end{equation}
where $a_{ij}=a_{i}-a_{j}$ and $\sigma_{ij}$ is the matrix which relates
the $g$ abelian differentials $v_{i}=z^{i-1}y(z)^{-1}dz$ with the 
canonical homology basis $\omega_{i}=\sum_{j}\sigma_{ij}v_{j}$.
In order to eliminate explicitely the $SL(2,{\bf C})$ freedom when choosing
the branch points on the sphere it is convenient to define the following 
harmonic ratios
\begin{equation}
\lambda_{i}=\frac{a_{i4}a_{56}}{a_{i5}a_{46}} \hspace*{1cm} i=1,2,3.
\label{h}
\end{equation}
Modular transformations now act on $\lambda_{i}$; for example under a Dehn 
twist along $A_{2}$ we have $T_{2}:\lambda_{i}\rightarrow \lambda_{i}/(
\lambda_{i}-1)$.

The computation of the higher genus cosmological 
constant for the heterotic string has been a
rather controversial issue. In what follows we will briefly review
the main problems found in such computations and the main 
features of the expression found by Knizhnik in  \cite{knizhnik} for the
genus-two cosmological constant of heterotic strings. 

While evaluating the functional integral for 
a heterotic string over a genus-two Riemann surface 
the main problem comes from the integration 
over the fermionic part of the supermoduli, i.e., the zero modes of 
the worldsheet gravitino. The two-loop cosmological constant in general
can be written as an integral over the supermoduli $m^{I}$ of an integrand
which factorizes into a holomorphic and antiholomorphic part (with respect the
$m^{I}$). However after integration over the fermionic moduli this 
factorization property is in general destroyed; using bosonization
\cite{martinec} is is argued that the integration over the gravitino zero
modes is equivalent to the insertion of $2g-2$ ($g\geq 2$) Picture Changing 
Operators (PCOs) whose correlation function destroys the holomorphic 
factorization of the original expression. Using this, Knizhnik proposed the
following expression for the genus-two cosmological constant of the 
ten-dimensional heterotic string
\begin{equation}
Z_{g=2}=\sum_{e,f,g} C(e,f,g) \int\prod_{i=1}^{6}\frac{1}{dv^{2}_{pr}}
T^{-5}\prod_{k<l}^{6} \overline{a_{kl}}^{-3}a_{kl}^{-2} 
\overline{\cal O}_{f}^{2}\overline{\cal O}_{g}^{2}[{\cal P}^{X}+
{\cal P}^{gh}_{e}]{\cal O}_{e},
\label{k}
\end{equation}
where $(e,f,g)$ are even spin structures, $C(e,f,g)$ are the phases dictated
by the GSO projection in the different heterotic string models; ${\cal O}_{e}=
(\det \sigma)^{2}\theta^{4}[e]$ are the partition functions for each set
of eight world-sheet fermions, matter and gauge, $T$ is given by
\begin{equation}
T=\int d^{2}z_{1}d^{2}z_{2} |(z_{1}-z_{2})y^{-1}(z_{1})y^{-1}(z_{2})|^{2}
\end{equation}
and ${\cal P}^{X}$, ${\cal P}_{e}^{gh}$ are respectively the matter and
ghost part of the correlator of two PCOs. Their explicit expressions are
\begin{equation}
{\cal P}^{X}=\frac{5}{8}a_{12}^{-1}\left[a_{23}a_{24}a_{25}a_{26}\frac{
P_{12}}{T}+a_{1}\leftrightarrow a_{2}\right],
\label{P10}
\end{equation}
with
\begin{equation}
P_{12}=\int d^{2}z_{1}d^{2}z_{2}\frac{(a_{1}-z_{1})(a_{1}-z_{2})}{
(a_{2}-z_{1})(a_{2}-z_{2})}\left|\frac{z_{1}-z_{2}}{y(z_{1})y(z_{2})}
\right|^{2}
\end{equation}
and
\begin{equation}
{\cal P}_{e}^{gh}=\frac{1}{4}a_{12}^{-1}\sum_{i=1}^{3}(a_{1}-A_{3}^{e})
(a_{1}-B_{i}^{e})(a_{1}-B_{i+1}^{e})(a_{2}-B_{i+2}^{e}),
\label{gh1}
\end{equation}
when $e=(12A_{3}^{e}||B_{1}^{e}B_{2}^{e}B_{3}^{e})$ or
\begin{equation}
{\cal P}_{e}^{gh}=\frac{1}{4}a_{12}^{-1}(a_{1}-A_{2}^{e})(a_{1}-A_{3}^{e})
(a_{2}-B_{2}^{e})(a_{2}-B_{3}^{e})
\label{gh2}
\end{equation}
if $e=(1A_{2}^{e}A_{4}^{e}||2B_{2}^{e}B_{3}^{e})$. $dv_{pr}$ is just the
volume of the $SL(2,{\bf C})$ projective group
\begin{equation}
dv_{pr}^{2}=\frac{d^{2}a_{4}d^{2}a_{5}d^{2}a_{6}}{|a_{45}a_{46}a_{56}|^{2}}.
\end{equation}

To get (\ref{k}) it has been assumed that the two PCOs have been
located respectively at $a_{1}$ and $a_{2}$, as can be seen from the 
fact that their correlation function diverges when $a_{12}\rightarrow 0$.
This fact is actually behind the lack of modular invariance of (\ref{k})
since modular transformations interchange the branch points and 
then do not preserve the insertion points of the PCOs. 
 
This expression can be easily applied not only to the ten-dimensional
supersymmetric heterotic string but also to other models without supersymmetry
and/or compactified dimensions. This is done simply by taking different
choices for the $C(e,f,g)$ phases and/or adding new internal fermionic sectors.

\section*{Appendix C: Weierstrass Elliptic Function}
\appendix

\def\theequation{C.\arabic{equation}}
\setcounter{equation}{0}

In this third appendix we will collect some useful results about the 
Weierstrass elliptic function ${\cal P}(z|\omega_{1},\omega_{2})$ 
\cite{Wang-Guo}. 

\begin{figure}
\let\picturenaturalsize=N
\def\picsize{0.5in}
\def\picfilename{period.eps}
\ifx\nopictures Y\else{\ifx\epsfloaded Y\else\input epsf \fi
\let\epsfloaded=Y
\centerline{\ifx\picnaturalsize N\epsfxsize \picsize\fi
\epsfbox{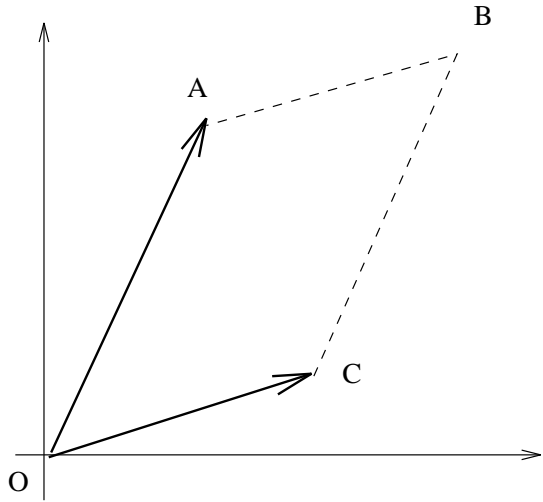}}}\fi
\caption{Fundamental paralelogram with sides $2\omega_{1}$ and 
$2\omega_{2}$.}
\label{periodf}
\end{figure}

A meromorphic function $f(z)$ is said to be an elliptic function if it is 
doubly periodic with semiperiods $\omega_{1},\omega_{2} \in {\bf C}$
\begin{equation}
f(z)=f(z+2m\omega_{1}+2n\omega_{2}),
\label{dp}
\end{equation}
with $m,n\in{\bf Z}$. Given (\ref{dp}) we see that $f(z)$ is determined
on the whole complex plane by its value in the fundamental 
paralelogram $OABC$ (fig. \ref{periodf}). 
Then as a corolary we see that any elliptic 
function without singularities in the fundamental paralelogram must
be a constant. In the same way it can be proven that the sum of the 
residues at the poles in $OABC$ vanishes.
The Weierstrass elliptic function ${\cal P}(z|\omega_{1},\omega_{2})$ 
is uniquely determined from the following three properties 
\begin{enumerate}
\item[-] ${\cal P}(z)$ is an elliptic function 
with a single pole located in $z=0$.
\item[-] Its principal part in $z=0$ is $1/z^{2}$.
\item[-] ${\cal P}(z)-z^{-2}$ tends to zero as $z\rightarrow 0$.
\end{enumerate}
It is defined by
\begin{equation}
{\cal P}(z|\omega_{1},\omega_{2})=\frac{1}{z^{2}}+
\sum_{m,n}{}^{'}\left[\frac{1}{(z-2m\omega_{1}-2n\omega_{2})^{2}}-
\frac{1}{(2m\omega_{1}+2n\omega_{2})^{2}}\right].
\end{equation}
Expanding in power series around $z=0$ we find
\begin{equation}
{\cal P}(z|\omega_{1},\omega_{2})=\frac{1}{z^{2}}+
\sum_{k=2}^{\infty} (2k-1)G_{2k} z^{2k-2},
\end{equation}
with $G_{2k}=\sum^{'}(2m\omega_{1}+2n\omega_{2})^{-2k}$.

An important result concerning elliptic functions is  that any 
elliptic function $f(z)$ can be written in terms of ${\cal P}(z)$ and
its derivative ${\cal
P}^{'}(z)$  which  is itself an elliptic function also. 
Suppose $f(z)$ is even, has a pole of order $2s$ at $z=0$ (or a zero for $s<0$)
and its remaining poles and zeroes in the fundamental paralelogram are located
respectively at $\{\beta_{1},\ldots,\beta_{k}\}$ and $\{\alpha_{1},\ldots,
\alpha_{k}\}$ then 
\begin{equation}
f(z)=C {\cal P}(z)^{s}\prod_{i=1}^{k} \frac{{\cal P}(z)-{\cal P}(\alpha_{i})}{
{\cal P}(z)-{\cal P}(\beta_{i})},
\label{c4}
\end{equation}
with $C$ a complex constant. If $f(z)$ is an odd elliptic function 
then $f(z)/{\cal P}^{'}(z)$ is even and  we can apply (\ref{c4}) and in the
case of a 
general $f(z)$ one can always  write it as the sum of an even and an odd 
piece. 

Since ${\cal P}^{'}(z)^{2}$ is an even elliptic function we know from what
we said in the last paragraph that it can be expressed in terms of 
${\cal P}(z)$. Locating the zeroes and the poles of ${\cal P}^{'}(z)$ 
we find
\begin{equation}
{\cal P}^{'}(z)^{2}= [{\cal P}(z)-e_{1}][{\cal P}(z)-e_{2}][
{\cal P}(z)-e_{3}],
\end{equation}
where $e_{i}$ 
can be written in terms of Jacobi's theta functions
\begin{eqnarray}
e_{1}&=&\frac{\pi^{2}}{12\omega_{1}^{2}}[\theta_{3}^{4}(0|\tau)
+\theta_{4}^{4}(0|\tau)], \nonumber \\
e_{2}&=&\frac{\pi^{2}}{12\omega_{1}^{2}}[\theta_{2}^{4}(0|\tau)
-\theta_{4}^{4}(0|\tau)], \nonumber \\
e_{3}&=&-\frac{\pi^{2}}{12\omega_{1}^{2}}[\theta_{2}^{4}(0|\tau)
+\theta_{3}^{4}(0|\tau)].
\end{eqnarray}
After some algebra it can be proven that 
\begin{equation}
{\cal P}(z|\omega_{1},\omega_{2})-e_{k}=
\left[\frac{\theta^{'}_{1}(0|\tau)}{2\omega_{1}\theta_{k+1}(0|\tau)}
\frac{\theta_{k+1}(v|\tau)}{\theta_{1}(v|\tau)}\right]^{2},
\label{apB}
\end{equation}
where $k=1,2,3$ and $v=z/(2\omega_{1})$. 

To finish, let us define the Weierstrass $\zeta$-function 
\begin{equation}
\int_{0}^{z} dz\left[{\cal P}(z)-\frac{1}{z^{2}}\right]=
-\zeta(z)+\frac{1}{z}.
\end{equation}
Taking derivatives with respect to $z$ in both sides of the last
formula we have
\begin{equation}
{\cal P}(z)=-\partial_{z}\zeta(z);
\end{equation}
$\zeta(z)$ is an odd function of $z$ but it is not an elliptic function, 
since it can be easily checked that
\begin{equation}
\zeta(z+2m\omega_{1}+2n\omega_{2})=\zeta(z)+2m\eta_{1}+2n\eta_{2};
\end{equation}
where $\eta_{1}$ and $\eta_{2}$ are related to the semiperiods by
\begin{equation}
\eta_{1}\omega_{2}-\eta_{2}\omega_{1}=\frac{i\pi}{2}.
\end{equation}

\end{document}